\documentclass[crop]{CSL}

\jname{International Journal of Astrobiology}

\usepackage{amsmath} 
\usepackage{graphicx}
\usepackage{multirow}
\usepackage{booktabs}
\usepackage{float}
\usepackage{hyperref} 
\usepackage{graphics} 
\usepackage{subfig} 
\usepackage{xcolor}
\usepackage{enumitem}
\usepackage{amsmath}
\usepackage{bm}
\usepackage{mathtools}
\usepackage{soul}

\usepackage[authoryear]{natbib}

\newcommand{\dd}{\mathrm{d}}
\newcommand{\eqn}[1]{Eq.\,\ref{#1}}

\newcommand{\fig}[1]{Fig.\,\ref{#1}}

\definecolor{orange}{rgb}{1,0.5,0}
\definecolor{sred}{rgb}{.5,0,0}

\renewcommand{\textbf}[1]{#1}

\begin{document}

\supertitle{Research Paper}

\title[Habitability of FFP's exomoons]{Presence of liquid water during the evolution of exomoons orbiting \textbf{ejected} free-floating planets}

\author[Roccetti \textit{et al.}]{Giulia Roccetti$^{1,2}$, Tommaso Grassi$^3$, Barbara Ercolano$^{2,4}$, Karan Molaverdikhani$^{2,4}$, Aur\'elien Crida$^5$, Dieter Braun$^{6}$, Andrea Chiavassa$^{5,7}$}

\address{\add{1}{European Southern Observatory, Karl-Schwarzschild-Str. 2, D-85748 Garching bei M\"unchen, Germany}; \add{2}{Universit\"ats-Sternwarte, Fakult\"at f\"ur Physik, Ludwig-Maximilians-Universit\"at M\"unchen, Scheinerstr. 1, D-81679 M\"unchen, Germany}; \add{3}{Max-Planck-Institut f\"ur extraterrestrische Physik, Giessenbachstr. 1, D-85748 Garching, Germany}; \add{4}{Exzellenzcluster ‘Origins’, Boltzmannstr. 2, D-85748 Garching, Germany};
\textbf{\add{5}{Universit{\'e} C{\^o}te d'Azur, Observatoire de la C{\^o}te d'Azur, CNRS, Laboratoire Lagrange, France};  \add{6}{Department of Physics, Center for Nanoscience
Ludwig-Maximilians-University of Munich
Geschwister-Scholl Platz 1, 80539 Munich, Germany}; \add{7}{Max Planck Institute for Astrophysics, Karl-Schwarzschild-Str. 1, 85748 Garching, Germany}}}

\corres{\name{Giulia Roccetti} \email{giulia.roccetti@eso.org}}

\begin{abstract}

Free-floating planets \textbf{(FFPs)} can result from dynamical scattering processes happening in the first few million years of a planetary system's life. Several models predict the possibility, for these isolated planetary-mass objects, to retain exomoons after their ejection. The tidal heating mechanism and the presence of an atmosphere with a relatively high optical thickness may support the formation and maintenance of oceans of liquid water on the surface of these satellites. In order to study the timescales over which liquid water can be maintained, we perform dynamical simulations of the ejection process and infer the resulting statistics of the population of surviving exomoons around free-floating planets. The subsequent tidal evolution of the moons' orbital parameters is a pivotal step to determine when the orbits will circularize, with a consequential decay of the tidal heating. We find that close-in ($a \lesssim 25$~R$_{\rm J}$) \textbf{Earth-mass moons} with CO$_2$-dominated atmospheres could retain liquid water on their surfaces for long timescales, depending on the mass of the atmospheric envelope and the surface pressure assumed. \textbf{Massive atmospheres are needed to trap the heat produced by tidal friction that makes these moons habitable.} For Earth-like pressure conditions ($p_0$ = 1~bar), satellites could sustain liquid water on their surfaces up to \textbf{52\,Myr}. For higher surface pressures (10 and 100~bar), moons \textbf{could be habitable} up to \textbf{276~Myr} and \textbf{1.6}~Gyr, respectively. \textbf{Close-in satellites experience habitable conditions for long timescales, and during the ejection of the FFP remain bound with the escaping planet, being less affected by the close encounter.} 

\end{abstract}
\keywords{atmospheres, habitability, tidal heating, planets and satellites.}

\selfcitation{Roccetti G \textit{et~al.} (2023) Presence of liquid water during the evolution of exomoons orbiting \textbf{ejected} free-floating planets 1--11. https://doi.org/xxxxx}

\received{xx xxxx xxxx}

\revised{xx xxxx xxxx}

\accepted{xx xxxx xxxx}

\maketitle


\section{Introduction}
The discovery of multiple free-floating planets (FFPs) \citep[e.g.,][]{zapatero2000,liu2013,luhman2014,liu2016} and FFP candidates \citep[e.g.,][]{sumi2011,bennet2014,henderson2016,mroz2020} is changing our understanding of the early evolution of planetary systems and planet formation theories. The largest population of FFPs discovered so far was found in the Upper Scorpius and Ophiuchus star-forming region by \citet{miret-roig2022}, containing between 70 and 170 FFPs, depending on the age assumed for the region. Those isolated planetary-mass objects can form in isolation, from a scaled-down version of a star formation process, through OB stars' wind erosion of a prestellar core, as aborted stellar embryos ejected from a stellar nursery, or from the ejection of a planet in a young planetary system due to close encounters between other giant planets, fly-by stars, or FFPs. The latter seems more common than previously thought: the number of FFPs found in the Upper Scorpius and Ophiuchus star-forming region is up to 7~times larger than the number of expected FFPs based only on core-collapse models. \textbf{{\citet{miret-roig2022}}} suggests that ejection due to dynamical instabilities could be very common during the first 10~Myr of a system's life (i.e., the upper limit of the age of their sample).

Close encounters between young giant planets are considered important to explain the high eccentricities and inclinations found in the populations of hot Jupiters \citep{beauge+nesvorny2011}. \citet{rasio+ford1996} and \citet{chatterjee2008} proposed a possible link between planetary systems hosting hot Jupiters and the formation of FFPs. During a close encounter, 
the dynamical scattering events can lead to the ejection of one or more giant planets from the \Fpagebreak planetary system, while the perturber can be excited from its initial orbit, reaching a hot  Jupiters  configuration after tidal flexing. 
This class of theoretical models predicts two FFPs per main sequence star in our Galaxy \citep{sumi2011}, i.e., around 2~M$_{\rm J}$ per star \citep{clanton+gaudi2015}. \textbf{\citet{barclay2017} shows that also rocky planets are potentially ejected from planetary systems due to scattering events with other giant planets, forming a population of rocky FFPs.} 

FFPs orbit around the Galactic centre and in star-forming regions, and although they are believed to be weakly radiated by nearby stars, they may present conditions favourable to sustain life. Such conditions could be met, for instance, by having an atmosphere rich in molecular hydrogen \citep{stevenson1999}, where the collision-induced absorption mechanism of molecular hydrogen (\cite{borysow2002}, \cite{frommhold2006}) could allow supporting liquid water on their surfaces. This condition could be maintained over 50~Gyr, depending on the core mass and the mass of the atmospheric envelope of the FFP \citep{mol-lous2022}.

As was already shown by \citet{rabago+steffen2019} and \citet{hong+et+al2018}, in the ejection scenario, a forming FFP can escape its planetary system retaining some of the moons formed in its circumplanetary disk. These dynamical scattering events strongly affect the orbital parameters of the \textbf{surviving} moons. 

As shown for example by \citet{heller-et-al2014}, the orbital parameters of the moons, together with the masses of the planet-moon system and the characteristics of the hosting star, are crucial for the maintenance of an atmosphere capable of retaining liquid water on the surface, even when these objects are outside the canonical stellar Habitable Zone (HZ). 

In the Solar System, Europa and Ganymede probably retain an ocean of liquid water beneath their surface \citep{spohn+schubert2002}, while for Callisto the results are less clear. The moon Io represents the most volcanically active object of our planetary system, as it is squeezed by the tidal torque of Jupiter acting on its interior. For Io and Europa the tidal heating mechanism represents the major energy source, while for Ganymede and Callisto, which reside on larger orbital configurations, radiogenic heating provides the largest fraction of the total energy budget \textbf{\citep{spohn+schubert2002}}. On the other hand, Saturn's moon Titan is the only satellite in the Solar System with a substantial atmosphere \textbf{\citep{hoerst2017}}. Although its formation history is still not fully understood, its dense atmosphere of nitrogen (94.2\%) and methane (5.7\%) \citep{catling+kasting2017} enables the moon to maintain liquid methane at its surface, which undergoes a methane cycle very similar to Earth's water cycle \textbf{\citep{lunine+atreya2008}}.

Even though moons orbiting exoplanets have yet to be detected, given the diversity of moons in the Solar System,  we expect them to be interesting objects both for benchmarking planet formation theories and for the search for biosignatures. \citet{heller2018} collected the proposed several candidates, based on the transit time variations (TTVs) technique \citep{teachey-et-al2018, teachey+kipping2018}, gravitational microlensing \citep{bennet2014,miyazaki2018},   and direct imaging \citep{lazzoni2022, ruffio2023}. As discussed by \citet{limbach2021}, FFPs are good candidates for the detection of exomoons. Without the glare produced by a nearby star, high-contrast imaging is not necessary to detect the photometric transit signal of a potential satellite and the detection of massive moons should be already possible with existing instrumentation. In particular, \citet{bachelet2022} propose a joint microlensing survey, using both ESA \textsc{Euclid} and NASA \textsc{Roman} missions, to detect and characterize FFPs and potentially discover the first exomoon orbiting around them.

These environments have been modelled by \citet{avila2021}, reproducing the chemical evolution of the atmosphere of an exomoon orbiting a FFP. They demonstrated the sustainability of liquid water on the exomoon's surfaces, by assuming a CO$_2$-dominated atmosphere, and tidal and radiogenic heating mechanisms as the main energy sources. Such a water reservoir could, in principle, assist the emergence of life.

Despite disagreements on the definition of habitability \citep{lammer-et-al2009}, and specific conditions needed for life to emerge \citep{jortner2006}, the presence of liquid water remains an essential criterion for an Earth-like habitable condition, thanks to water being an excellent solvent \citep{lubineau1999}. To this aim, in the Solar System, the oceans underneath the icy surfaces of Jupiter's and Saturn's moons are studied in search of habitable conditions \citep{hussmann2006}. Even on Earth, life can be found in very extreme environments dominated by liquid water, at temperature and pressure conditions very different from the Earth's surface (i.e. the depth of the oceans). Life adapted to live in these extreme conditions has developed different metabolisms, which are not based on photosynthesis \citep{irwin2020}. 

Life, at least as we know it, depends on informational polymers, either in the form of oligonucleotides or polymers of amino acids. While their formation and synthesis nowadays are driven by proteins in water, for its polymerization and especially for the phosphorylation of nucleotides, at least a partial dry state seems to be necessary \citep{powner2009,toner2020}. This means that moons of FFPs scenarios are especially interesting where surface water is in touch with the atmosphere or where hydrothermal systems occur near the surfaces of the moon. Here the estimation that initial waters are about four orders of magnitude less abundant than on Earth is encouraging \textbf{\citep{avila2021}}, since early Earth was likely a rather water-dominated setting. Non-equilibrium drivings such as thermal gradients \citep{ianeselli2022jan,ianeselli2022march} from volcanic drivings such as on Io are interesting, not only to drive microscale water cycles in pores partially filled with water, but also for driving active systems such as heated surface ponds or geysers. All these settings could have provided an oscillatory system where molecules accumulate in the dry state and create polymers while they are from time to time exposed to water to perform the replication cycles for Darwinian evolution.

Hereinafter, for the purpose of this study, we will define habitable conditions referring only to the presence of liquid water on the surface of the moons, aware of the complex debate around this specific topic (e.g., \citealt{cockell2016}). Since the maintenance of liquid water is controlled by the evolution of the orbital parameters \citep{reynolds+cassen1978,scharf2006,henning2009,heller2012,heller+barnes2013}, a relatively rapid circularization of a moon's orbit will prevent long-term stability of tidal heating and hence reduce the timescale necessary for life to potentially emerge, assuming biological timescales comparable to what we measure on Earth \citep{pearce2018}.

Here, we present a detailed study of the orbital parameters' evolution of the FFP-moon system and its influence on the production and maintenance of liquid water, starting from dynamical simulations. We assume a Jupiter-like planet to be ejected from its planetary system due to close encounters with other giant planets. An initial population of moons is placed around the escaping planet to study the survival rate of the moons after the ejection process, and how the orbital parameters of the \textbf{surviving} moons are affected by the dynamical scattering event. 

The evolution of the orbital parameters of the \textbf{surviving} moons is obtained by integrating the differential equations from standard tidal models \citep{hut1981, bolmont2011}, where tides are generated both by the planet on the moon and vice versa. This temporal evolution, coupled with the modelling of the thermal properties of the optically thick atmosphere, allows us to calculate the surface temperature as a function of time, and therefore to  determine the best moon candidates to support the formation of liquid water, for a timescale compatible with the habitable conditions of the surface of the exomoons. 

The paper is organized as follows: in the second section, the numerical methods and modelling are described, in the third section we present the results, \textbf{in the fourth section we discuss the limitations of our models and results}, and in the last section the broader implications and the conclusions are discussed.


\section{Methods}\label{sec:methods}

\subsection{Formation of moons around a Jupiter-like planet}

Circumplanetary disks are thought to be the birthplaces of moons around gas giant planets, both considering the core accretion and the gravitational instability formation theories \citep[e.g.][]{canup2002, shabram2013, szulagyi2017}. \citet{cilibrasi2018} and \citet{cilibrasi2021} presented population synthesis works for the formation of satellites around a Jupiter-like planet. In their models, seeds can grow by accreting dust via streaming instability, and due to the influx of material from the protoplanetary disk, satellites can reach masses comparable to the Galilean system ones. However, some features of Saturn's moon system (i.e., the mass-distance relation of the regular moons, and the variation of the bulk composition of the satellites) cannot be explained by this formation mechanism. \citet{Charnoz+2011} have shown that Saturn's moons can rather originate from Saturn's rings, where solid particles can aggregate beyond the Roche radius of the planet, as self-gravity becomes stronger than tidal forces. \citet{Crida-Charnoz-2012} and \citet{crida+charnoz2014} generalised this satellite formation mechanism to most planets in the Solar System (and thus probably to other stellar systems). However, this process takes place after the planets are formed (and ejected), so that the satellites would not be affected by the dynamical history, in contrast to satellites formed inside the circumplanetary disk.

For the purpose of this study, we use the statistics of the formation of moons around a Jupiter-like planet in a circumplanetary disk from \citet{cilibrasi2021}. They performed N-body simulations of the planet-moons systems, taking into account the interaction between the satellites, which can migrate inward into the planet, be engulfed by the planet itself, collide with each other and be ejected by the system. \textbf{The final mass distribution of the formed satellites' systems shows that systems less massive than the Galilean one are more frequent, representing 85\% of their population, although previously \citet{cilibrasi2018} found the opposite result}. Since satellite seeds (with an initial mass of 10$^{-7}$\,M$_{\rm J}$) are continuously added to their simulations to replace lost satellites, the ones added near the end of the simulations had sensibly less time to interact, grow, and evolve and will bias the distribution towards low masses. This is not relevant in our dynamical simulations, where the satellites will be considered test particles. \textbf{The mass of the moons becomes relevant only in the subsequent tidal evolution and in determining the surface temperature. For this reason, it is possible to limit our analysis to the Earth-mass satellites, that are capable of trapping the heat generated by tidal friction, given their massive atmospheres. For completeness, the results obtained using the mass distribution from \citet{cilibrasi2021} are also shown. Given our setup, this represents a scenario with an increased probably of having relatively massive atmospheres around satellites  lighter than the Galilean moons, as it will be discussed later.} 

Circumplanetary disks cannot be considered isolated objects, since they are continuously accreting material from the hosting protoplanetary disk. The population synthesis approach is very dependent, among other things, on the protoplanetary disk's temperature profile, which greatly varies with the distance from the star, and influences the dynamics of protosatellites \citep[e.g., ][]{cilibrasi2021}. For that reason, the statistics of the formation of moons around a Jupiter-like planet cannot be used for other planets' mass and location in the protoplanetary disk.

Despite these limitations, we employ their final moons \textbf{orbital parameters population} (which is almost uniform in the logarithmic semi-major axis space up to $\sim 300$\,R$_{\rm J}$) as the input of our dynamical model, to analyse the effect of the interaction of giant planets and their ejection, on the distribution of the initial moons. 


\subsection*{Dynamical simulations}

In this work, dynamical simulations are performed to study the ejection process of a gas giant planet from a planetary system and to understand the influence of close-encounter events on the survivability of the moons. The impact of the dynamical scattering events on the orbital parameters of the moons is also investigated.

We consider a planetary system with a Sun-like star (1\,M$_{\odot}$) and three Jupiter-mass (1\,M$_{\rm J}$) planets. \textbf{This specific initial set-up and the planets' initial conditions are taken from \citet{rabago+steffen2019}, being their set of initial configurations proven to be capable of producing close encounters and be a reliable benchmark}. To perform the simulations, we use the N-body code \textsc{REBOUND} \citep{rein2012} with the IAS15 integrator \citep{everhart1985, rein2015}, which is an adaptive non-symplectic time step integrator up to the 15th order able to integrate planetary close-encounters for billions of orbits. 

The innermost planet is always placed at 5~AU (i.e., present-day Jupiter, conversely to \citealt{rabago+steffen2019} which used 3~AU), and the other two planets are randomly uniformly placed between 1.2 and 1.4~times the period of the previous one, reaching a maximum orbital distance for the third planet of less than 10~AU. Initial orbital eccentricities and inclinations of the planets are randomly generated from a Rayleigh distribution with a Rayleigh parameter of~0.01, while the mean anomaly, the longitude of ascending node, and the argument of pericenter, are randomly uniformly generated between 0 and $2\pi$. In total, we end up with $17$ free initial parameters for the planets, but, applying different machine learning classification algorithms \textbf{as Principal Component Analysis (PCA, to find dominant modes), t-distributed Stochastic Neighbourhood Embedding (t-SNE, which allows to include also non-linear relations),} and random forests from \textsc{scikit-learn} \citealt{pedregosa2011}), we do not find any correlation between the initial conditions and the outcome of the simulations. This confirms that the subsequent dynamical evolution is chaotic. 

To avoid collisions that could result in the subsequent merging of the planets, we select a minimum distance between the bodies during the encounters. As shown by \citet{li2020}, in hydrodynamical simulations of giant planet collisions, mass exchange and merging between the two planets strongly diminishes for an impact parameter larger than twice the diameter of the bodies. Therefore, we set 5~R$_{\rm J}$ as the minimum distance. If this distance is reached during the evolution, the simulation is stopped, and a new one is started. 

Dynamical simulations are evolved for a maximum of 10~Myr, \textbf{following the observational constraint in \citet{miret-roig2022}}. In our model, a planet is considered ejected from the system when it has (i) an orbital distance greater than 100~AU, and (ii) an orbital eccentricity larger than one. The first condition ensures that the planet is truly ejected far away from the system, i.e., where the interaction with the star is relatively weaker and allows the study of the planet-moons system as isolated. The second condition prevents a secular evolution of the ejected planet, as we are only interested in ejections due to a last strong close encounter event. Secular ejection of a giant planet can result in even more favourable \textbf{conditions} for the survivability of the moons \citep{hong+et+al2018}, but it would have required more computational time for the moon's evolution, making these calculations technically impractical. In this way, we obtain a lower bound to the estimation of the survivability rate of the satellites.

From all the simulations between the planets and the central star, we select some simulations with the ejection of the innermost planet as the final outcome. \textbf{Since in \citet{cilibrasi2021} the initial distributions of the orbital parameters of the moons are calculated only for a Jupiter-mass planet located at 5~AU, we assume the moons to orbit around the initially innermost planet.} In this way, we are implicitly assuming that other moons cannot be captured from the populations orbiting around the other two giant planets during close encounters. Our initial set consists of 26\,293 moons that are massless particles with initial semi-major axes, eccentricities, and inclinations with respect to the host planet as in \citet{cilibrasi2021}. To ensure the validity of the massless moons assumption, we calculated the Safronov number \citep{morbidelli2018} for an average moon as
\begin{equation}
    \Theta = \frac{v_{\rm esc}^2}{2\,v_{\rm orb}^2} \sim 10^{-1}\,,
\end{equation}
where $v_{\rm esc}$ is the escape velocity of the moon with respect to \textbf{another moon}, and $v_{\rm orb}$ is the orbital velocity of the moon. A Safronov number smaller than 1 indicates that the velocity acquired after the dynamical scattering between two moons is not sufficient for a moon to escape the potential well of the planet, and thus, a typical value of the order of $10^{-1}$ confirms a weak gravitational interaction among the moons themselves.

The evolution of the moons is relatively computationally expensive, and therefore, to reduce the calculation time, we avoid placing the moons around the Jupiter-like planet from the very beginning of the simulation. In fact, given a simulation where the first planet is ejected, the moons are placed 20~years before the last dynamical scattering event, \textbf{to let every moon to complete at least half an orbit before the close encounter.} \textbf{To do so, we first select the simulations with the ejection of the innermost planet, and we store the time of the last close encounter. Then, we rerun these simulations starting from 20~years before the last close encounter, but this time including the moons, that are now evolved until the planet reaches 100~AU from the host star.}
Earlier close encounters between the planets are likely to happen in the simulations, and they will probably affect the initial distributions of semi-major axes, eccentricities, and inclinations of the moons, while also leading to the escape of some outer moons. However, the outcome of the simulations is mainly affected by the impact parameter of the last close encounter \textbf{(i.e., the minimum distance between the two planets during the close encounter event), as shown by \citet{hong+et+al2018}}. Hence, we only consider the last close encounter in our simulations.


\subsection*{Tidal model} \label{sec2}

The tidal evolution of the orbital parameters (i.e., semi-major axis and eccentricity) of the moons is important for determining the timescale in which habitable conditions can be found on the surface of the moons. To this aim, we solve the time-dependent tidal heating differential equations to study the evolution of the orbital parameters. \citet{hut1981} proposed the first comprehensive set of coupled differential equations for a constant time lag model, which takes into account the spin evolution of both the primary and secondary bodies, while only tides generated on the primary body from the secondary are considered. Since we are interested in the tides generated on the moon (secondary body), we use an updated version developed by \citet{bolmont2011} that includes spin evolution, and tides generated on both bodies. The model is built to study the evolution of the tidal interaction between brown dwarfs and planets orbiting around them, and given the similarity of the system (i.e., absence of a central star) it is suitable to be employed for our case study, where the primary body is the Jupiter-mass FFP.

The coupled differential equations, which consider the secular evolution of the semi-major axis $a$, the eccentricity of the orbit $e$, the spin frequency of the planet $\Omega_p$, and of the moon $\Omega_m$, are
\begin{align}
    \frac{1}{a} \frac{\dd a}{\dd t} = &- \frac{1}{T_{\rm m}} [N_{\rm a1}(e) - \frac{\Omega_{\rm m}}{n} N_{\rm a2}(e) ] \nonumber\\ 
    &- \frac{1}{T_{\rm p}} [N_{\rm a1}(e) - \frac{\Omega_{\rm p}}{n} N_{\rm a2}(e) ]
    \label{eq:a}
\end{align}

\begin{align}
    \frac{1}{e} \frac{\dd e}{\dd t} = &- \frac{9}{2 T_{\rm m}} [N_{\rm e1}(e) - \frac{11 \Omega_{\rm m}}{18 n} N_{\rm e2}(e) ] \nonumber \\
    &- \frac{9}{2 T_{\rm p}} [N_{\rm e1}(e) - \frac{11 \Omega_{\rm p}}{18 n} N_{\rm e2}(e) ]
    \label{eq:e}
\end{align} 

\begin{align}
    \frac{1}{\Omega_{\rm m}} \frac{\dd \Omega_{\rm m}}{\dd t} = - \frac{\gamma_{\rm m}}{2 T_{\rm m}} [N_{\rm o1}(e) - \frac{\Omega_{\rm m}}{n} N_{\rm o2}(e)] 
    \label{eq:omega_m}
\end{align} 

\begin{align}
    \frac{1}{\Omega_{\rm p}} \frac{\dd \Omega_{\rm p}}{\dd t} = &- \frac{\gamma_{\rm p}}{2 T_{\rm p}} [N_{\rm o1}(e) - 
    \frac{\Omega_{\rm p}}{n} N_{\rm o2}(e)] \nonumber \\ 
    &- \frac{2}{R_{\rm p}} \frac{\dd R_{\rm p}}{\dd t} - \frac{1}{r_{\rm g}} \frac{\dd r_{\rm g}}{\dd t}\,,
    \label{eq:omega_p}
\end{align} 
where $n$ is the mean motion, $\gamma = \frac{h}{I \Omega}$ is the ratio between the orbital angular momentum ($h$) and the spin angular momentum ($I \Omega$), $r_{\rm g}$ is the radius of gyration of the planet from $I_{\rm p} = M_{\rm p} (r_{\rm g} R_{\rm p})^2$, and $N_{\rm a}$, $N_{\rm e}$, and $N_{\rm o}$ are polynomial functions of the eccentricity (see Appendix~A). The dissipation timescale of the planet is calculated as 
\begin{equation}
     T_{\rm p} = \frac{1}{9} \frac{M_{\rm p}}{M_{\rm m} (M_{\rm p} + M_{\rm m})} \frac{a^8}{R_{\rm p}^{10}}\frac{1}{\sigma_{\rm p}},
\end{equation}
while $T_{\rm m}$ is the dissipation timescale of the moon (obtained by swapping the $p$ and $m$ subscripts in the previous equation). The internal dissipation factor \textbf{of the FFP} $\sigma_p$ is 2.006 $10^{-60}$~g$^{-1}$~cm$^{-2}$~s$^{-1}$ \textbf{from \citet{bolmont2011}, while the moon has}
\begin{equation}
   \sigma_{\rm m} = \frac{2}{3} \frac{G k_{\rm 2} \Delta t_\ell}{R^5_{\rm m}}\,,
\end{equation}
where $G$ is the gravitational constant, $k_{\rm 2}$ is the Love number characteristic of the moon, and $\Delta t_\ell$ is the constant time lag. \textbf{In a constant time lag model, the tidal bulge lags behind or ahead of the moon with a fixed time difference. \citet{efroimsky+lainey2007} defines the relation between the quality factor of the moon $Q$ and the angular distance $\delta$ as}
\begin{equation}
    Q^{-1} \sim 2 \delta = 2 \Omega_{\rm m}^{\rm rev} \Delta t_\ell \,,
\end{equation}
\textbf{where $\Omega_{\rm m}^{\rm rev}$ is the Keplerian velocity of the moon around the FFP, which depends on the orbital parameters of the FFP-moon system. It follows that}
\begin{equation}
   \sigma_{\rm m} = \frac{k_{\rm 2}}{Q} \frac{G}{3 R^5_{\rm m} \Omega_{\rm m}^{\rm rev}},
\end{equation}
\textbf{with the factor $k_{\rm 2}/Q$ playing a crucial role in determining the dissipation timescale of the moons. The tidal evolution is discussed hereinafter both for Earth-mass satellites, as well as for moons folowing the mass distribution of \citet{cilibrasi2021}. This is possible because moons are assumed to be massless particles in the dynamical simulations, and thus the orbital parameters at this stage are not affected by the mass.}

\textbf{For Earth-mass satellites, we assume $k_{\rm 2}$ = 0.302 and $Q$ = 280 from \citet{lainey2016}. In the case of moons with the mass distribution of \citet{cilibrasi2021}, following \citet{henning2009}, we use $Q$ = 100 and}
\begin{equation}
 k_{\rm 2} = \frac{3}{2}\frac{1}{1+\bar{\mu}}\,,
\label{k2}
\end{equation}
where the effective rigidity of the body is
\begin{equation}
\bar{\mu} = \frac{19}{2}\frac{\mu}{\rho_{\rm m} g R_{\rm m}}\,,
\label{mu}
\end{equation}
with $g$ the surface gravity of the satellite, $\mu$ the material rigidity, and $\rho_{\rm m}$ the mass density of the satellite, which we assume to be the average mass density of the four moons of the Galilean system ($\rho_{\rm m} = 2.58$~g\,cm$^{-3}$). \citet{yoder+peale1981} estimated a value of $\mu = 5 \times 10^{11}$~dyne~cm$^{-2}$ for a common rocky body.

The integration of the equations is performed using \textsc{solve\_ivp} from the \textsc{Python} library \textsc{scipy} \citep{virtanen2020}, using the Backward Differentiation Formula (BDF) solver with adaptive time step, and absolute and relative tolerances of $10^{-40}$ and  $10^{-12}$, respectively. For each planet-moon system, the initial conditions of $a_0$ and $e_0$, with $t_0 = 1$~Myr, are taken by the final semi-major axes and eccentricities from the dynamical simulations, while the initial angular frequencies of the planet and the moon require a more detailed discussion. Since from our tests we noticed that the initial $\Omega_{\rm m,0}$ does not noticeably affect the results of the evolution, we assume that all the moons at the beginning are tidally locked with the planet, allowing us to calculate an initial spin frequency of the moon that depends on its orbital distance. Regarding the Jupiter-like planet, as an initial condition, we assume the same spin frequency as Jupiter's current one
\begin{equation}
\Omega_{p,0} = \frac{2\pi}{P_{\rm J}} = 1.78 \cdot 10^{-4}~{\rm rad\,s}^{-1},
\end{equation}
where \textbf{$P_{\rm J}$} is the rotational period of Jupiter.

For the evolution of the angular frequency of the planet, we find that the evolution of the radius as a function of time plays the dominant role. Since a FFP does not experience stellar irradiation, we cannot use radius evolution models for gas giant planets around a star. \citet{leconte2011} developed a model for determining the radius and other properties of isolated objects (i.e. brown dwarfs, FFPs) from their masses and ages. The main difference between FFPs and irradiated objects resides in the direct bloating of the outer atmospheric layers of the latter, which slows down the gravitational and thermal evolution of the planet itself. When using this model for the evolution of the planetary radius, we need to set up our time integration extremes between 1~Myr and 10~Gyr. Following our assumption of an escaping FFP with a Jupiter mass, the initial radius of the FFP is set at 1.6 R$_{\rm J}$ at 1~Myr, and it shrinks to 1~R$_{\rm J}$ at 10~Gyr (see Appendix~B for the planetary radius evolution profile).

Lastly, moons that enter the Roche radius of their host planet are removed from our sample as tidal forces disrupt them. The Roche radius for a rocky satellite can be calculated as
\begin{equation}
    d_{\text{Roche}} = 2.44 \cdot R_{\rm p} \left(\frac{\rho_{\rm p}}{\rho_{\rm m}} \right)^{\frac{1}{3}} = 
    \begin{cases}
    2.43~{\rm R_J} & \text{if } R_{\rm p} = 1.6~{\rm R_J}\\
    1.52~{\rm R_J} & \text{if } R_{\rm p} = 1.0~{\rm R_J}\,,
    \end{cases}
    \label{eq:roche}
\end{equation}
from \citet{roche1849}, where $\rho_{\rm p}$ is the mass density of Jupiter, $R_{\rm p}$ is the giant planet radius, and $\rho_{\rm m}$ is the mass density of the satellites. \textbf{The values in Eq.~\ref{eq:roche} are calculated for Earth-mass satellites (using the Earth mass density $\rho_{\rm m}$ = 5.51~g~cm$^{-3}$), and for the \citet{cilibrasi2021}'s mass distribution (with $\rho_{\rm m} = 2.58$~g\,cm$^{-3}$, as previously discussed). Since the mass density of the satellites is constant,} the Roche radius is the same for all the moons, and is compared with their perihelion distance $d_{\rm per} = a (1 - e)$, to check whether the moons, at their closest distances to the planet, are inside the Roche radius.

\subsection*{Atmospheric modelling}

Without considering the irradiation from the star, in our model we only have one possible energy source, i.e., the tidal heating mechanism. Other energy sources can also play important roles under specific conditions, which usually depend on the formation history of the moon. Incident planetary radiation \citep{haqq-misra+heller2018, dobos-et-al2017}, secular cooling after the moon's formation, radiogenic heating due to the decay of active radionuclides \citep{nimmo2020}, and runaway greenhouse effect \citep{heller+barnes2015} are other possible heating mechanisms. In addition to the thermal budget, the optical depth of a potential atmosphere could trap part of the emitting radiation producing a considerable increase in the temperature of the planet, allowing the possible formation and maintenance of liquid water \citep{avila2021}. In addition to this, without the stellar heliosphere, a FFP and its moons are considerably less shielded from galactic cosmic rays, which represent the main chemical driver for the formation of water.
Given the fact that the link between the composition and density of Solar System moons' atmospheres and their formation histories is still not well understood \citep[e.g., ][]{lammer2018}, we study a suit of atmospheric conditions by setting different surface pressures on each \textbf{surviving} moon. Considering only tidal heating, we are calculating a lower estimate of the possible moons with habitable conditions at their surfaces, since other energy sources (which are highly sensitive to the planet-moons' initial conditions and formation history) could also play major roles.  

The effective temperature of the moon is

\begin{equation}
    T_{\rm eff}^4 = \frac{\dot{E}_{\rm tidal}}{4 \pi \sigma_{\rm sb} \epsilon_{\rm r} R_{\rm m}^2}\,,
\end{equation}

where $\sigma_{\rm sb}$ is the Stefan-Boltzmann constant, $\epsilon_r=0.9$ is the infrared emissivity factor \citep{henning2009}, $R_{\rm m}$ is the radius of the moon, and the tidal heating energy flux is

\begin{equation}
    \dot{E}_{\rm tidal} = \frac{21}{2} \frac{G k_2 M_{\rm p}^2 R_{\rm m}^5 n e^2}{Q a^6}\,,
    \label{eq:tidal}
\end{equation}

where $G$ is the gravitational constant, $n$ the mean motion of the orbit, $a$ the semi-major axis, $e$ the eccentricity, and $M_{\rm p} = 1$~M$_{\rm J}$ the mass of the FFP. The second-order Love number $k_2$ and the quality factor $Q$ of the moons depend on the internal structure and composition of the satellites, \textbf{and they were introduced in the previous section.} We note that the tidal heating energy flux follows a power-law relation between $a$, $e$, and $M_{\rm m}$. In particular, in \eqn{eq:tidal}, expanding the mean motion $n$ in terms of the semi-major axis and the radius of the satellite $R_m$ in terms of the mass, we obtain
\begin{equation}
    \dot{E}_{\rm tidal} \propto  a^{-\frac{15}{2}} e^2\hspace{3pt} M_{\rm m}^{\frac{5}{3}}\hspace{3pt}\,,
    \label{eq:tidal_power_law}
\end{equation}
which shows that the semi-major axis plays the dominant role in the tidal heating mechanism, followed by eccentricity.

Considering the possible presence of an atmosphere, different atmospheric escape processes can prevent the atmosphere from remaining stable. Many features of the system, including the mass and the exospheric temperature of the moon, the magnetic field of the planet, the incident flux of cosmic rays, and the outgassing processes of the mantle, determine the mass and the composition of the atmospheric envelope and their evolution in time. Without modelling the interior of the moon and the other mentioned processes, we assume a CO$_2$-dominated atmosphere, and we only consider thermal (hydrostatic) escape as a criterion to assign an atmospheric envelope to the different moons. Following \citet{seager2010}, therein Eq.\,4.50, we compare the molecular thermal velocity with the gravitational escape velocity
\begin{equation}
    \frac{M_{\rm m} ^ {\frac{2}{3}}}{T (\tau = 0)} > \frac{36 k_{\rm B}}{G \mu_{\rm m} m_{\rm H}} \left(\frac{3}{4 \pi \rho_{\rm m}} \right) ^ {\frac{1}{3}}\,,
\label{atm_escape}
\end{equation}
where $M_{\rm m}$ is the mass of the satellite, \textbf{$k_{\rm B}$ is the Boltzmann constant}, $\mu_{\rm m} = 44.01$ is the mean molecular weight of a CO$_2$ molecule, $m_{\rm H}$ is the mass of a hydrogen particle, and the exospheric temperature \textbf{, T($\tau = 0$),} is calculated as the temperature at the top of the atmosphere, where the optical depth $\tau$ is equal to zero, i.e.,
\begin{equation}
    T (\tau = 0) = 2^{- \frac{1}{4}} T_{\rm eff}\,.
\end{equation}

\textbf{The maximum altitude of the atmosphere for each moon is calculated fixing the pressure at the top of the atmosphere to $p (\tau = 0) = 10 ^ {-5}$~bar, and taking into account the difference in the effective temperature.} Combining the definition of scale height $H = {k_{\rm B} T}{\,\left(\mu_{\rm m} m_{\rm H} g\right)^{-1}}$ \textbf{(with $g$ the surface gravity of the moon)} and the hydrostatic equilibrium, we obtain the maximum altitude
\begin{equation}
    z_{\rm max} = - H[T(\tau = 0)] \ln\left[\frac{p(\tau = 0)}{p_0}\right]\,,
    \label{zmax}
\end{equation} 
where $p_0$ is the surface pressure, which in our study is \textbf{a parameter varied between $0.1$ and $100$~bar. Note that, in the case of varying masses of the moons (following \citet{cilibrasi2021}), the surface gravity of the satellites will also affect the scale height and thus the maximum altitude of the atmosphere.}

\textbf{Different atmospheric configurations are determined by the formation history and the evolution of the moons. For this reason, the mass is not the only parameter to have a direct influence on the atmospheric surface pressure \citep{lammer2018}, as shown for example by comparing the Earth ($p$ = 1\,p$_\oplus$, $m$ = 1\,M$_\oplus$), Venus (91\,p$_\oplus$, 0.82\,M$_\oplus$), and Titan (1.5\,p$_\oplus$, 0.02\,M$_\oplus$). Therefore, we decided to investigate also various surface pressure values in the Solar System, assuming Venus as the upper limit. However, for the moons in \citet{cilibrasi2021}'s sample that are smaller than the ones in the Galilean system, $p_0=$ 100~bar might be considered unphysical, according to what we observe in the Solar System.}

To calculate the surface temperature, we assume the atmosphere to have both a radiative and a convective regime. We consider a 1D vertical model in which the pressure is always calculated assuming hydrostatic equilibrium, while the temperature at each layer is derived depending on the atmospheric heat transport regime
\begin{equation}
 T=
    \begin{cases}
     T_{\rm eff}\left[\frac{1}{2} (1+D\tau)\right]^{1/4} & \text{if } \frac{\dd\log T}{\dd\log p} < \nabla_{\rm ad}\\ \\
    T_{\rm b}\left(\frac{p}{p_{\rm b}}\right)^{\nabla_{\rm ad}}  & \text{if } \frac{\dd\log T}{\dd\log p} > \nabla_{\rm ad}\,,
    \end{cases}
\label{eq:thermal_profile}
\end{equation}
where $D = 1.5$ \citep{marley2015} is the diffusivity factor, $T_{\rm b}$ and $p_{\rm b}$ are the temperature and pressure estimated at the boundary between the radiative and the convective regimes, while $\nabla_{\rm ad}$ is the adiabatic lapse rate of the atmosphere, which depends on the adiabatic index \textbf{$\Gamma$} ($\nabla_{\rm ad} = \frac{\Gamma - 1}{\Gamma}$).

The surface temperature of the moon is determined by the optical depth
\begin{equation}
    \tau(p) = - \int_{z|p}^{z_{\rm max}} k_{\rm r}(z) \rho_{\rm a} dz = \int_{p}^{p(\tau = 0)} k_{\rm r}(p) g^{-1} \dd p\,,
\end{equation}
where $\rho_{\rm a}$ is the mass density of the atmosphere, $p(\tau = 0) = 10^{-5}$~bar is the pressure at the top of the atmosphere, and $k_{\rm r}$ is Rosseland mean opacity at each layer. The Rosseland mean opacities for a CO$_2$-dominated atmosphere are taken from the tables in \citet{badescu2010}, specifically computed for a FFP case. 

\section{Results}\label{sect:results}

\subsection{Dynamical simulations}

We perform 8\,000 simulations, including only the star with its planetary system. After 10~Myr, only a few planetary systems remain stable, while the majority become dynamically unstable, causing one planet to be ejected from the planetary system or a collision among two of the giant planets. The three giant planets in the simulations are observed to often exchange their relative positions. As a result, we find a comparable rate of ejection among the three planets. The outcomes are shown in Table \ref{tab1}, where we note that the distribution of the ejections between the three planets is roughly the same, and it is relatively common. Such a high rate of ejections is expected since we adopted \citet{rabago+steffen2019} model, which is designed to produce ejections among giant planets.

\begin{table}[h]
    \tabcolsep4pt
    \processtable{Outcomes of the dynamical simulations. \label{tab1}}
    {\begin{tabular}{@{\hspace*{4pt}}lcc@{\hspace*{4pt}}}
    \rowcolor{Theadcolor}
        Outcome & Occurrences & Percentage \\\hline
        No ejection & 43 & 0.54\%\\\hline
        Planet \#1 ejected & 1402 & 17.53\% \\\hline
        Planet \#2 ejected & 1508 & 18.85\%\\\hline
        Planet \#3 ejected & 1438 & 17.97\%\\\hline
        Collision & 3609 &  45.11\%\\\hline
    \end{tabular}}{\begin{tablenotes}
    \end{tablenotes}}
\end{table}

\begin{figure}
    \centering
    \includegraphics[width=0.48\textwidth]{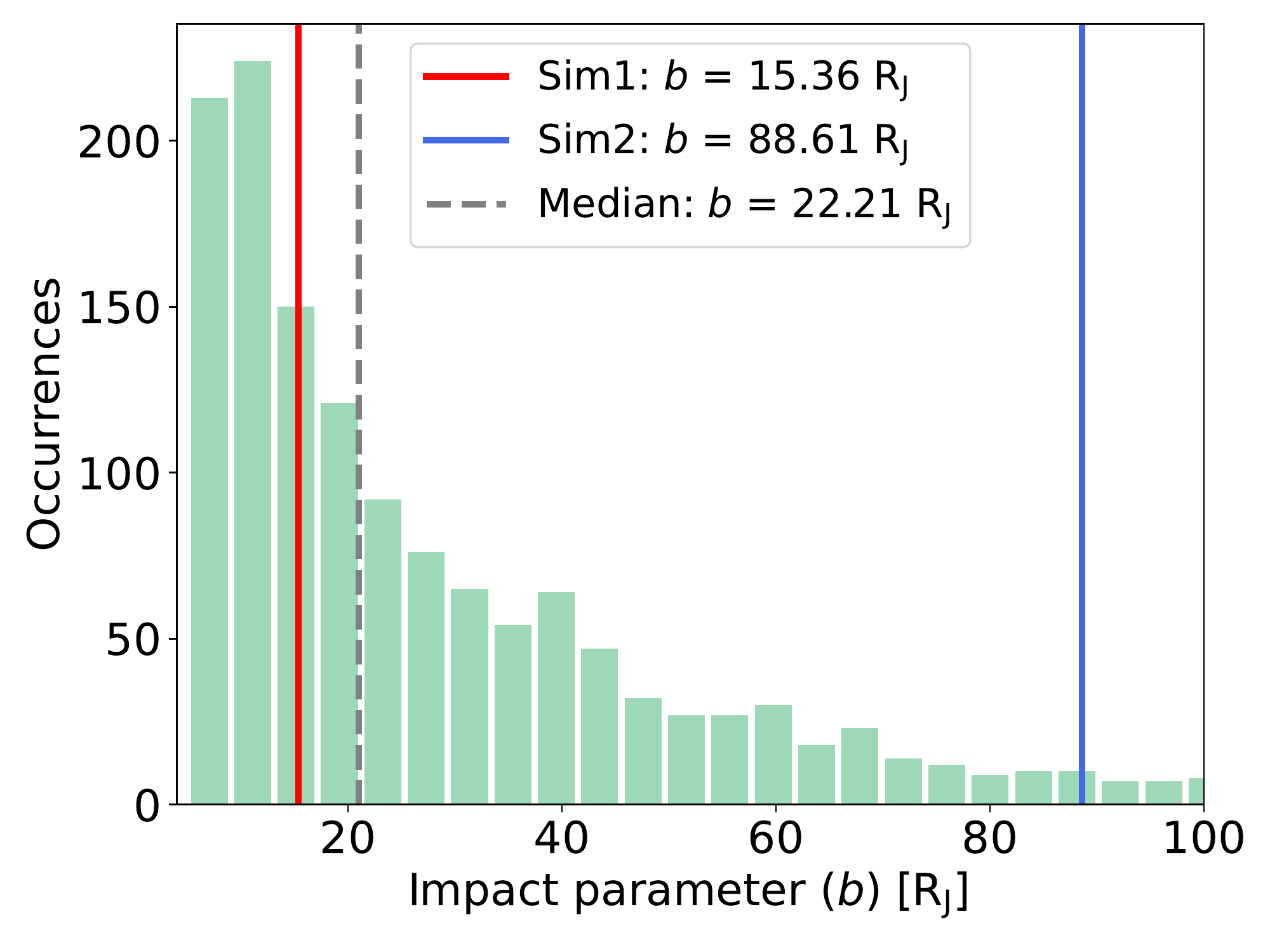}
    \caption{Minimum close encounter distance among all the dynamical scattering events for all the 1\,402 simulations with the ejection of planet \#1 as the outcome. The simulations are performed between three Jupiter-mass planets and a Sun-like star. For impact parameters smaller than 5\,${\rm R_J}$, we consider that the two planets collided with each other, representing a different outcome of the simulations. The plot is cut at 100~${\rm R_J}$, while there is still a very long tail of simulations with larger impact parameters. In red and blue, we show the impact parameters of the last close encounter of Sim1 and Sim2, which will be analysed in more detail during this work. We note that $b_1 = 15.36~{\rm R_J}$ for Sim1 is below the median of the distribution ($b_{\rm med} = 21.02~{\rm R_J}$), while $b_2 = 88.61~{\rm R_J}$ is well above the median. Note that while we show the impact parameters of the closest dynamical scattering event for all the 1\,402 simulations, for Sim1 and Sim2 we show the impact parameter of only the last close encounter, when the moons are placed around the planet.}
    \label{fig:impact_par}
\end{figure}

The 26\,293 moons from \citet{cilibrasi2021} are placed around the ejected planet in the simulations where the ejection of planet \#1 is the final outcome. \citet{hong+et+al2018} found that the impact parameter of the close encounter event plays a major role in determining the survivability of the moons. To this aim, we present the results of two simulations with different impact parameters. Sim1 and Sim2 are two simulations with respectively smaller ($b_1 = 15.36~{\rm R_J}$) and much larger ($b_2 = 88.61~{\rm R_J}$) impact parameters than the median of the distribution ($b_{\rm med} = 21.02~{\rm R_J}$), as shown in Fig. \ref{fig:impact_par}. Sim1 is in the second quartile of the distribution, while Sim2 is in the last one.

\begin{figure*}
  \includegraphics[width=\textwidth]{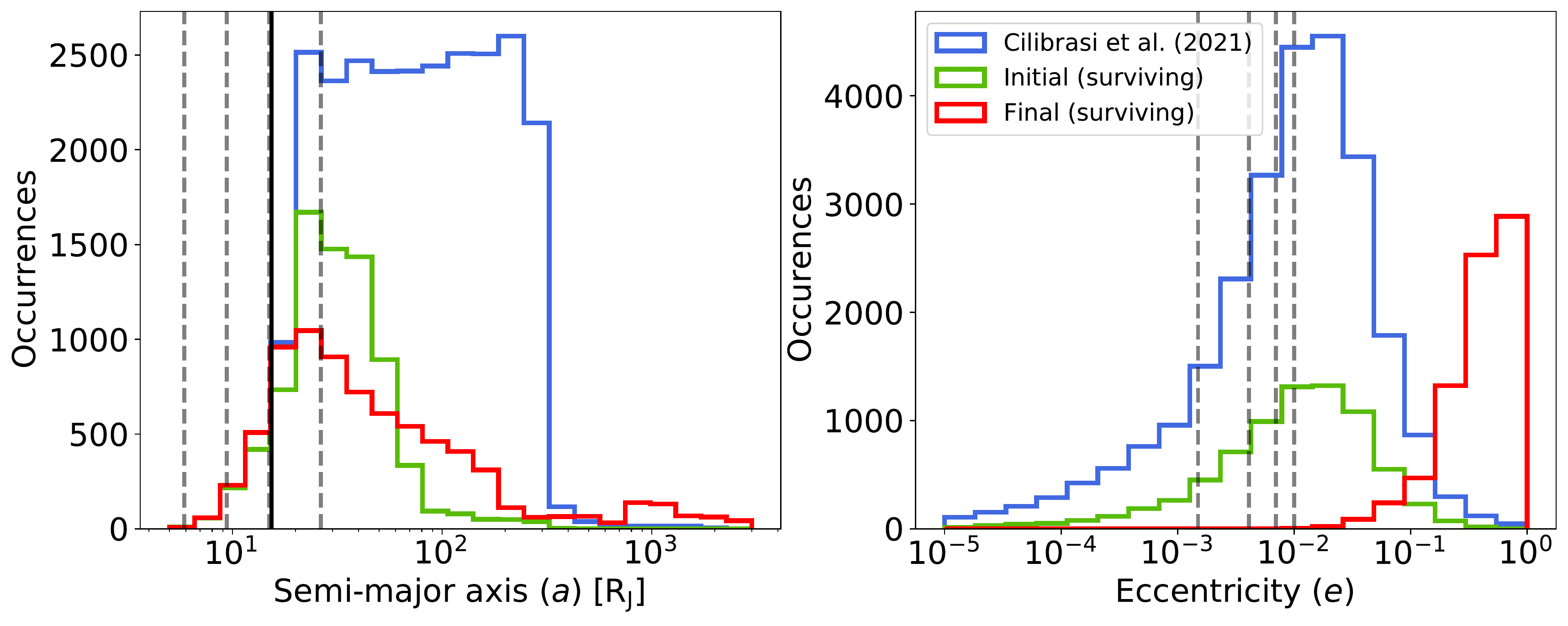}
  \caption{Dynamical evolution and survivability of the moons in Sim1 (impact parameter $b_1$ = 15.36 $R_{\rm J}$). We compare the distributions of semi-major axes (left panel) and eccentricities (right panel) between the initial statistics of the total 26293 moons from \citet{cilibrasi2021} (in blue), the initial distributions of the \textbf{surviving} moons (in green) and the final distributions of the \textbf{surviving} moons after the close encounter event (in red). Note that the green distribution is a subset of the total initial distribution in blue. In Sim1, 7570 moons remained bound with the planet after the close encounter, corresponding to a 28.79\% of survivability rate. The final distribution of the semi-major axis of the \textbf{surviving} moons is much more spread out, while the final eccentricities substantially increase due to the dynamical scattering event. Grey dashed lines represent the Galilean moons orbital parameters, while the solid black line is placed at the last close encounter impact parameter.}
  \label{fig:sim1}
\end{figure*}

\begin{figure*}
  \includegraphics[width=\textwidth]{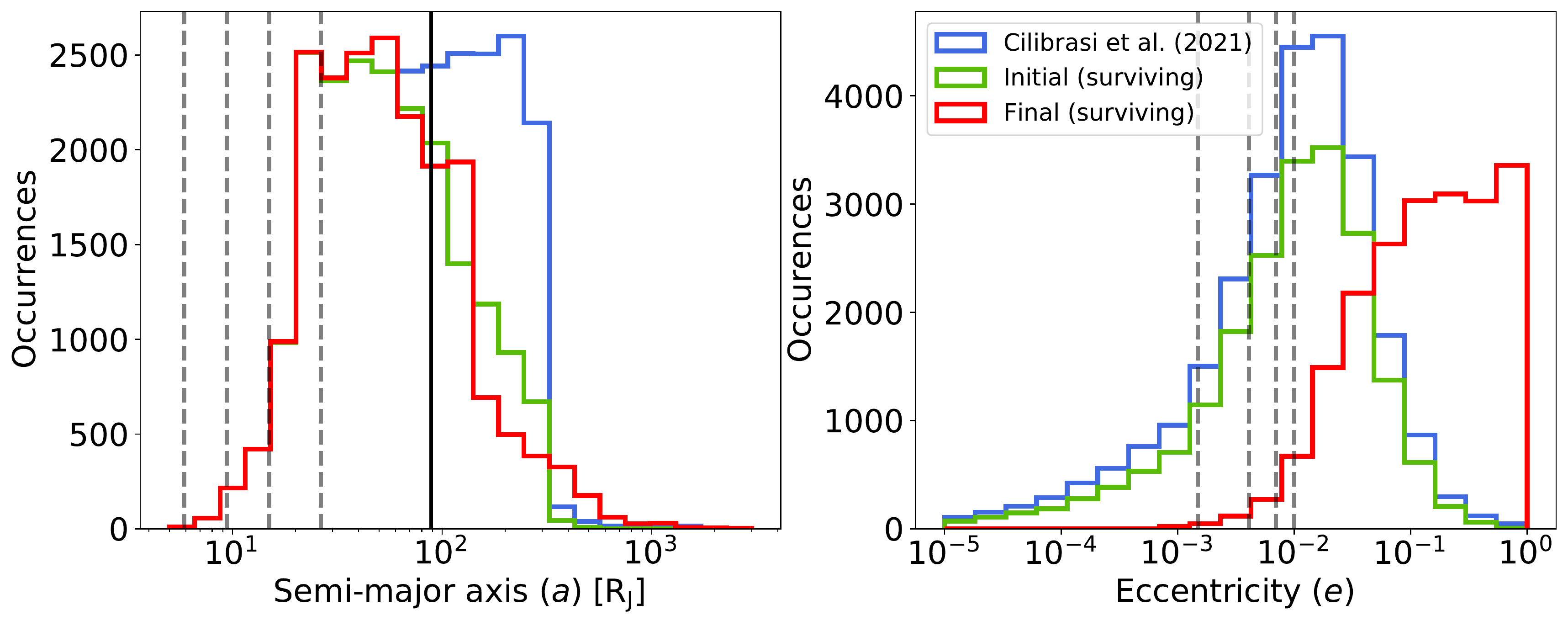}
  \caption{Dynamical evolution and survivability of the moons in Sim2 (impact parameter $b_2$ = 88.61~$R_J$). We compare the distributions of semi-major axes (left panel) and eccentricities (right panel) between the initial statistics of the total 26293 moons from \citet{cilibrasi2021} (in blue), the initial distributions of the \textbf{surviving} moons (in green) and the final distributions of the \textbf{surviving} moons after the close encounter event (in red). Note that the green distribution is a subset of the total initial distribution in blue. In Sim2, 19945 moons remained bound with the planet after the close encounter, corresponding to a 75.87\% of survivability rate. Having a larger impact parameter, moons in close-in orbits (i.e. with semi-major axes compared to the Galilean system ones) are less affected by the dynamical scattering event and less perturbed. Outer moons experience again an increase in eccentricity and semi-major axis. Grey dashed lines represent the Galilean moons orbital parameters, while the solid black line is placed at the last close encounter impact parameter.}
  \label{fig:sim2}
\end{figure*}

Moons are considered to have survived if they remain inside the Hill radius of the host planet at the end of the simulation. Since the Hill radius depends on the orbital distance of the planet from the central star, at 100~AU the Hill radius for a Jupiter-mass planet is 6.83~AU (cf. 0.34~AU of Jupiter at 5~AU), allowing the ejected planet to retain also relatively eccentric and distant moons. We find that moons initially closer to the planet have a higher probability of remaining bound. The initial spatial configuration of the planets (i.e., their eccentricity, inclination, mean anomaly, longitude of ascending node, and argument of pericenter) does not influence the statistics of the \textbf{planet's} simulations, \textbf{as tested by PCA and t-SNE algorithms}. Also, only the initial \textbf{semi-major axis} of the moons plays a role in the final distribution of the surviving ones, not their exact position on the orbit.

In Figs~\ref{fig:sim1} and \ref{fig:sim2} we show the moons' orbital parameters evolution due to the last close encounter events happening in Sim1 and Sim2. In blue, the semi-major axis and eccentricity's distributions of the initial population of moons from \citet{cilibrasi2021} are shown. In green, the initial orbital parameters of the fraction of moons which survived the ejection process are represented, i.e, they are a subset of the initial populations in blue. Lastly, in red, we show the final semi-major axes and eccentricities of the \textbf{surviving} moons, after they evolved in time until the planet is considered ejected. We notice that moons with relatively small initial semi-major axes have higher chances of surviving the close encounter event, as described by the difference between the blue and green populations. This is in accordance with the findings of \mbox{\citet{rabago+steffen2019}}. Regarding the initial eccentricities, they do not play \textbf{a role} in selecting the final \textbf{surviving} moons.

After the close encounter event, the final orbital parameters distributions (in red) are substantially affected by the gravitational interaction with the perturber. This translates to a wider distribution of the final semi-major axes of the \textbf{surviving} moons and a steep increase in the final distribution of the eccentricities.

Comparing Figs.\,\ref{fig:sim1} and~\ref{fig:sim2}, we notice that the moons' survival rate greatly varies with the impact parameter of the simulation; in Sim1 28.79\% of the moons survived the close encounter event, while 75.87\% of moons survived in Sim2. 

In the simulation with the lower impact parameter (Sim1), most of the moons with semi-major axis smaller than Ganymede's (14.97~${\rm R_J}$) remain bound to the planet, while in Sim2 this boundary is at a larger orbital distance,  larger than, e.g., Callisto's (26.33~${\rm R_J}$). In particular, the semi-major axis of the innermost moons in Sim2 is not affected by the dynamical event. Regarding the increase in the orbital eccentricities, in Sim1 they usually reach values larger than $0.1$, whereas in Sim2 the limit is much lower, at around $0.01$. This again suggests that the innermost moons are less gravitationally affected by the perturber, and so their orbits remain more stable.

The final distributions of the semi-major axis and eccentricity of Sim1 are the starting point for the orbital parameters' evolution, as described in the Methods. Hereinafter, we only use the results of Sim1, being the lower estimate on the number of \textbf{surviving} moons. \textbf{In our analysis, the moons will be considered Earth-mass during the tidal evolution stage, but we will compare the results using the mass distribution of \citet{cilibrasi2021}.}

\textbf{In the latter case, the final distribution of the masses of the surviving moons is a subset of the initial distribution. The mass of the moons does not affect the ejection process, with the survival rate only depending on the initial semi-major axis. Since the mass distribution weakly correlates with the semi-major axis in \citet[][Fig. 10]{cilibrasi2021}, the final mass distribution resembles the initial one. However, the ejection process favours moons with $a\lesssim 50~{\rm R_J}$ (see Fig~\ref{fig:sim1}), which tend to have lower masses. In particular, we find that around 8\% of the surviving moons are more massive than Europa, with the most massive moon of the sample being slightly less massive than the Earth.}


\subsection{Tidal evolution}

The \textbf{surviving} moons in Sim1 are evolved with the tidal model (Eqs.\,\ref{eq:a}, \ref{eq:e}, \ref{eq:omega_m} and, \ref{eq:omega_p}). First, we notice that \textbf{625} moons reached, during the \textbf{tidal} evolution, orbital distances smaller than the Roche radius of the Jupiter-like planet, which is \textbf{2.43}~R$_{\rm J}$ for a planetary radius of 1.60~R$_{\rm J}$ (Eq.\,\ref{eq:roche}). Removing them, we end up with a final population of \textbf{6\,945} \textbf{surviving} moons. Moons disrupted by tidal friction form debris material, which accumulates in rings. We expect that FFPs can retain rings of debris material around them, like Saturn and Jupiter's ones, and following the mechanism proposed by \citet{Crida-Charnoz-2012, crida+charnoz2014} new satellites can also generate around FFPs forming from their rings, in a similar way as \citet{VanLieshout+2018} showed that this mechanism might form second-generation planets around white dwarves. We do not study these second-generation moons here.

\begin{figure*}
  \centering
  \includegraphics[width=\textwidth]{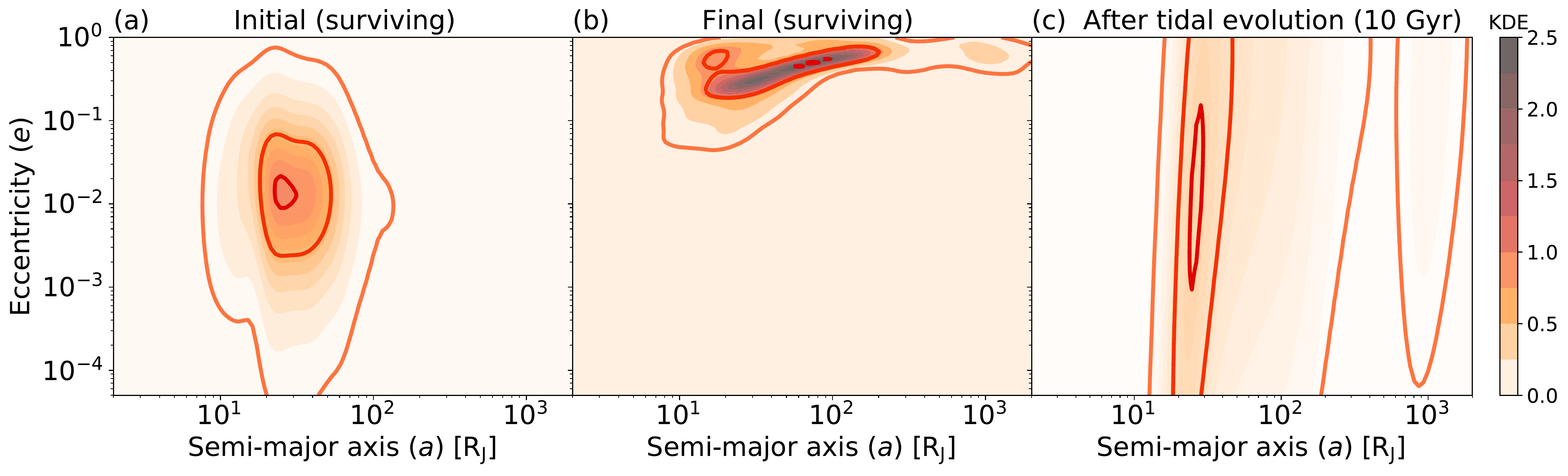}
  \caption{Kernel Density Estimation (KDE) of the correlation between the semi-major axes and eccentricities of the \textbf{surviving} moons. On the left panel (a), we note a flat distribution of the initial shape of the KDE. In the central panel (b) the correlation after the ejection process happened in Sim1 between the semi-major axis and eccentricity appears, and the entire population of \textbf{surviving} moons is shifted at higher eccentricities, as already observed in Fig. \ref{fig:sim1}. In panel (c), the final correlation after 10~Gyr of tidal evolution for the \textbf{Earth-mass surviving} moons becomes \textbf{much} more dispersed than the previous distribution (panel b). The red contour lines show the 5th, 50th, and 95th percentiles of the distributions. The normalization in each panel is constrained as $\int\int KDE(u, v)\,\dd u\, \dd v = 1$, where $u=\log(a)$ and $v=\log(e)$.}
  \label{KDE}
\end{figure*}

In Fig. \ref{KDE} we show the correlations as the Kernel Density Estimation \citep[i.e., a Gaussian-kernel-based probability density, ][]{Scott1992} between the eccentricity and semi-major axis for (a) initial orbital elements of the moons who survived the last close encounter (green distribution in Fig. \ref{fig:sim1}), (b) after the ejection process (red distribution in Fig. \ref{fig:sim1}, removing the moons inside the Roche radius), and (c) after their orbital parameters' evolution at 10~Gyr \textbf{(considering Earth-mass moons)}. Between panel (a) and panel (b), the general trend observed is the same as in Fig. \ref{fig:sim1}, with a steep increase in the eccentricities. In panel (b), the new orbital parameters show some degree of power-law correlation, absent in panel (a), which is a result of the dynamical scattering event. In panel (c), this correlation \textbf{vanishes} due to the tidal evolution of the eccentricities and semi-major axes of the satellites, as the orbits circularize due to tidal dissipation. In other words, it is possible to interpret each event as a filtering operator that shapes the probability density function in two opposite directions; while the first operator (a)$\to$(b) considerably reduces the dispersion in the ($a$, $e$) parameter space (mainly in the $e$ component), the second (b)$\to$(c) broadens the probability density function in \textbf{the eccentricity variable}. As expected, \textbf{both processes} play a crucial role in shaping the probability of finding a moon with given orbital characteristics.

We also note that the tidal evolution of the moons (b)$\to$(c) strongly depends on its initial semi-major axis and eccentricity. \textbf{Moons tend to migrate inward, reaching orbital configurations closer to the FFP. For these satellites, which experience more tidal heating, a more marked decay of the orbital eccentricity is observed. Conversely, more distant satellites have a weaker interaction with the planet, making their eccentricities more stable. The timescale for the circularization of the orbit depends mostly on the dissipation factor of the moon} ($\sigma_{\rm m}$)\textbf{, as shown in} \citet{bolmont2011}.


\subsection{Surface Temperature}

The surface temperature of all the \textbf{6\,945} moons is calculated assuming different surface pressure values ($p_0$ = \textbf{0.1~bar,} 1~bar, 10~bar and 100~bar) for a CO$_2$-dominated atmosphere. Considering hydrostatic atmospheric escape, over the entire surface temperature evolution, moons which get particularly close to the FFP are not considered capable of retaining an atmosphere, in accordance with Eq.\,\ref{atm_escape}, and their number varies according to the different surface pressure conditions. After this last stage, the remaining moons constitute the final investigated population. 

The thermal profile of all the moons at every time step is computed using Eq. \ref{eq:thermal_profile}, between 1~Myr and 10~Gyr. During the evolution, the orbital parameters of the moons change due to tidal dissipation, and $\rm T_{eff}$  changes consequently. The surface temperature of the moons is influenced not only by the evolution of the effective temperature, but also by the surface pressure, and hence, by the total mass of the atmosphere. In particular, the total mass of the atmospheric envelope of each satellite varies as its maximum altitude changes as a function of the effective temperature (Eq. \ref{zmax}).

\begin{figure}
    \centering
    \includegraphics[width=0.45\textwidth]{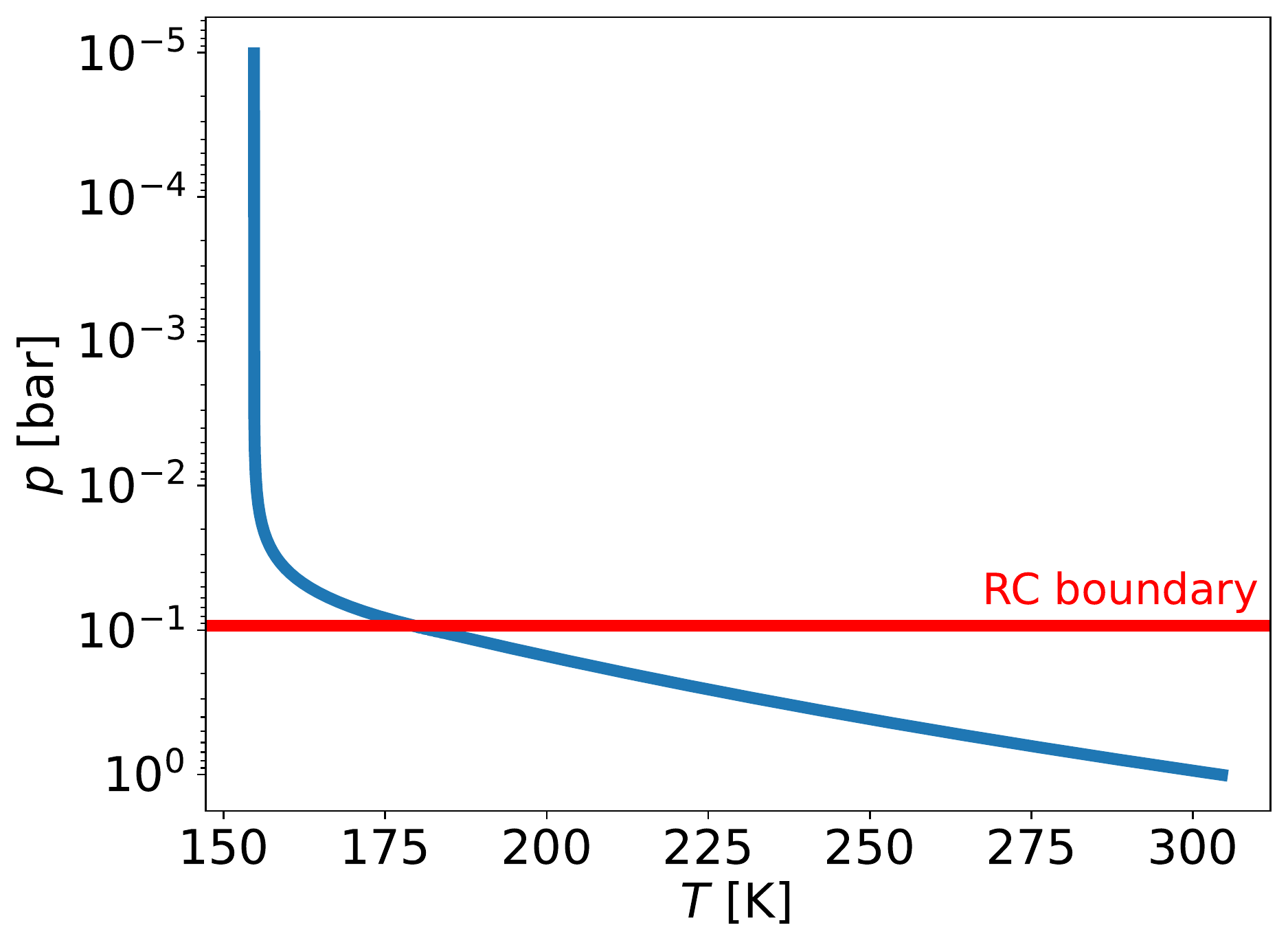}
    \caption{\textbf{Earth-mass} moon's atmosphere temperature profile, indicating the boundary between the radiative and convective regimes (red line). Above the boundary, the atmosphere is in the radiative regime, while below the boundary, the heat transport is dominated by the convective motion of the air. Habitable conditions ($T_{\rm surf}$ = 305~K) are reached for this particular moon: as initial conditions we consider the $p_0$ = 1~bar pressure case, $T_{\rm eff}$ = 183.9~K, and surface gravity $g$ = 981~cm\,s$^{-2}$ which lead to an atmosphere scale height of 2.96~km and air density at the top of the atmosphere of $10^ {-8}$~g~cm$^{-3}$.}
    \label{radiative_vs_convective}
\end{figure}

In Fig.\,\ref{radiative_vs_convective} we show one example of a \textbf{Earth-mass} moon with $T_{\rm eff}$ = 183.9~K, a surface gravity of 981~cm\,s$^{-2}$, a scale height of $H$ = 2.96~km, and a maximum height of the atmosphere of $z_{\rm max}$ = 34.1~km. The boundary between the radiative and the convective regime (red horizontal line) is at $\sim 10^{-1}$~bar, followed by an increase of temperature in the convective domain, which, in this particular case, results in a temperature at the surface compatible with the presence of liquid water.

\begin{figure*}[h]
    \centering
    \includegraphics[width=\textwidth]{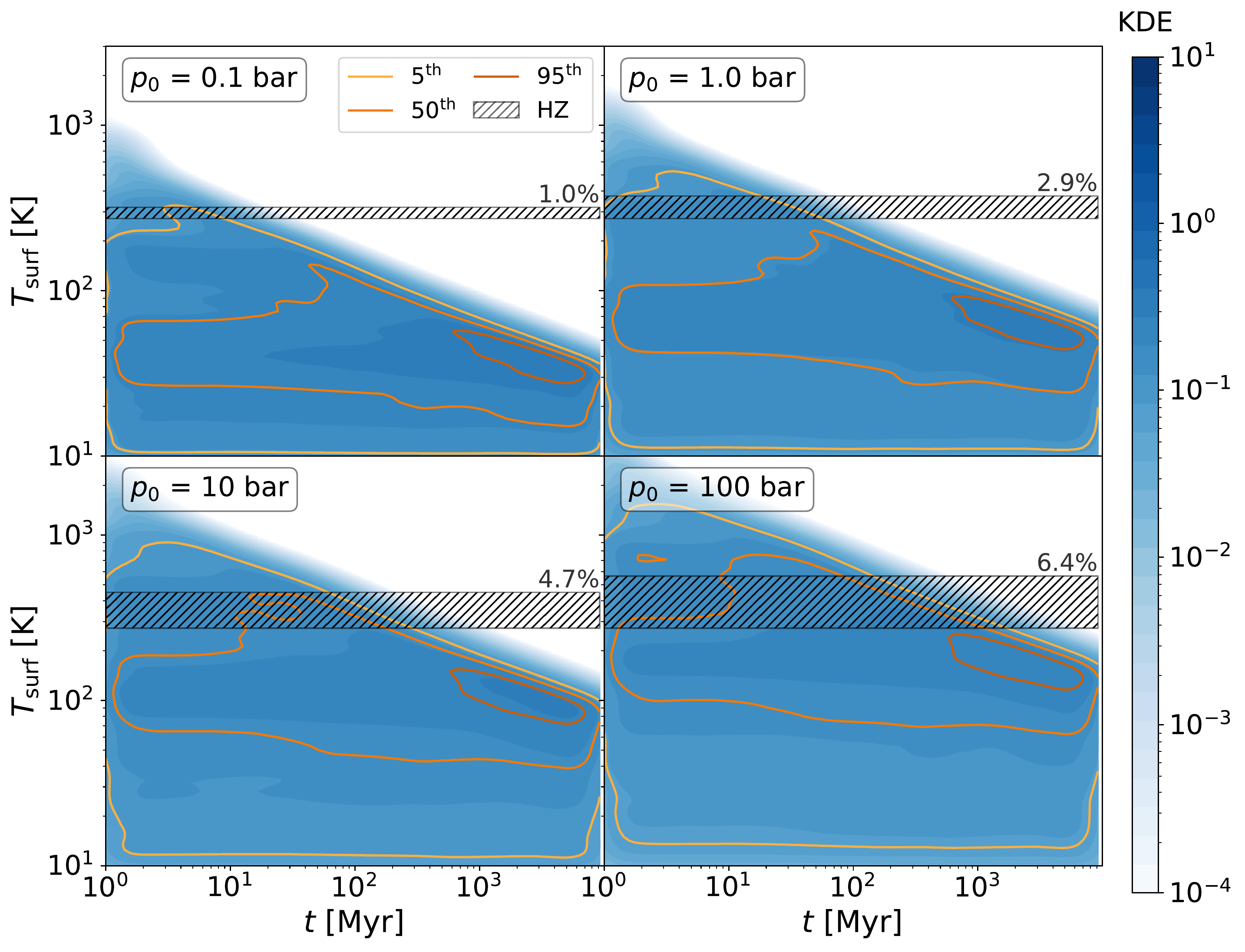}
    \caption{Probability density function, calculated as the KDE, to find a moon at a certain surface temperature as a function of time. \textbf{Different} panels show increasing surface pressures. The hatched area represents the HZ, and the contour lines show the 5th, 50th, and 95th percentiles of the distributions. We note that the presence of more massive and substantial atmospheres increases the surface temperature of the moons and the number of moons inside the HZ. Increasing $p_0$ also increases the timescale spent in the HZ. Above the HZ areas, we report the probability for a moon orbiting a FFP to lie in the HZ during its lifetime. The normalization of the KDE is analogous to \fig{KDE}.}
    \label{pdf}
\end{figure*}

\begin{figure}
    \centering
    \includegraphics[width=0.50\textwidth]{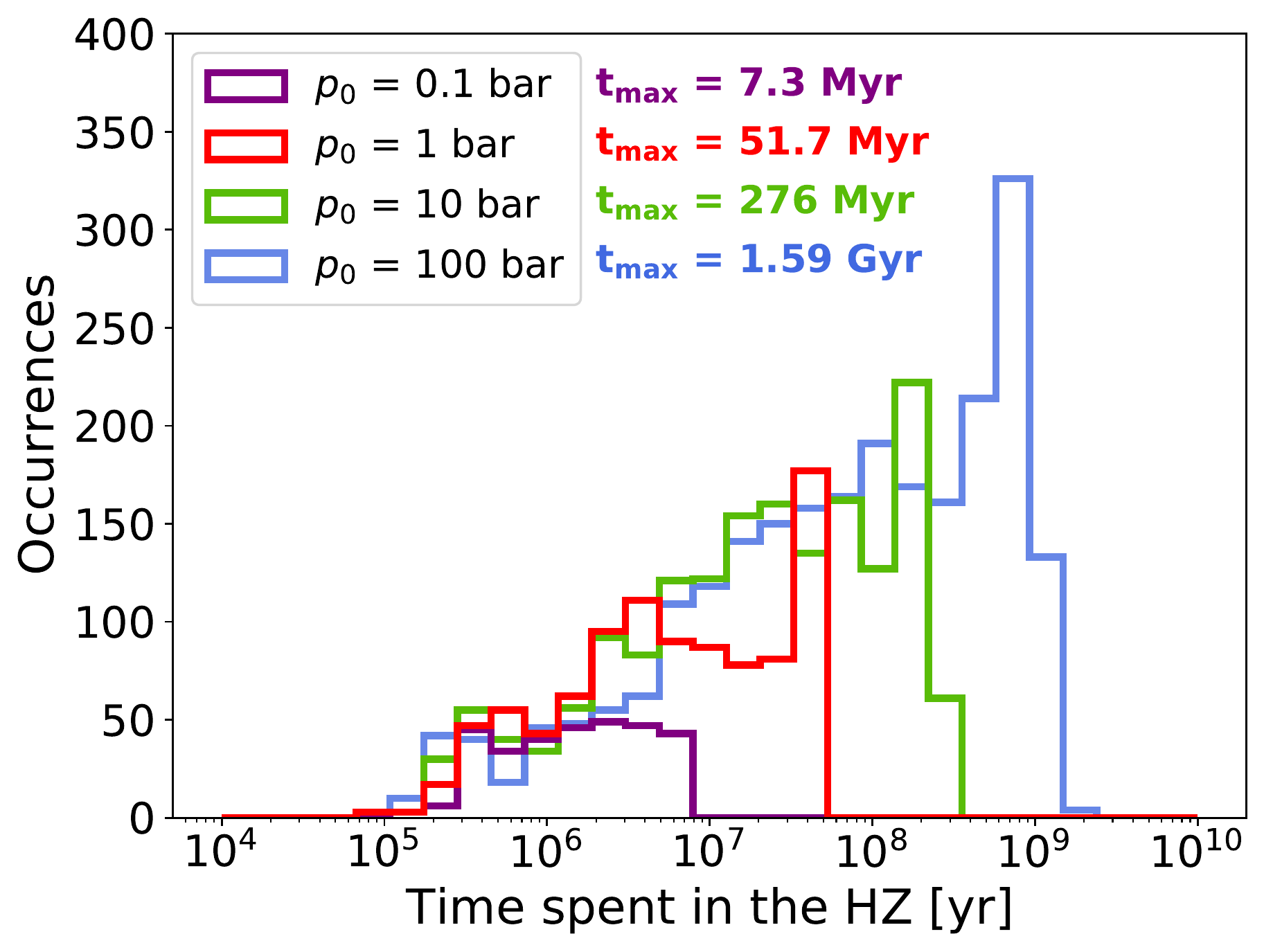}
    \caption{Time spent in the HZ for the moons which survived the close-encounter event of Sim1. Note that moons with a more substantial atmosphere ($p_0$ = 100~bar) can retain liquid water on their surface up to \textbf{1.6~Gyr}. For $p_0 = 0.1$~bar, moons could be habitable up to 7.3~Myr, while for an Earth-like surface pressure ($p_0$ = 1~bar) liquid water could be retained up to \textbf{52~Myr} on the surface, and for $p_0$ = 10~bar liquid water could be retained up to \textbf{276~Myr.}}
    \label{final_plot}
\end{figure}

The evolution of the surface temperature of the moons greatly varies with different surface pressure conditions \textbf{(from $p_0$ = 0.1~bar to $p_0$ = 100~bar)}, as shown in Fig.\,\ref{pdf}. Instead of representing each moon, we compute the probability density function (PDF), calculated with a KDE, of finding one moon in a certain time ($t$) and temperature ($T_{\rm surf}$) interval, indicated by the colour bar. For the sake of clarity, the plot is limited to $T_{\rm surf}>10$~K, i.e., moons colder than this limit are not shown, and the KDE is calculated without considering them. Colder moons are removed because their surface temperature could be dominated by other heating mechanisms (e.g., radiogenic heating). 
We note that by increasing the surface pressure ($p_0$) and thus the mass of the atmospheric envelope, satellites reach higher surface temperatures and increasingly populate the HZ \textbf{(i.e., the region in the parameter space where moons get surface temperatures compatible with the presence of liquid water)}, as reported by the number in each panel that indicates the fraction of the moons in the HZ. The temperature boundaries of the HZ get wider increasing $p_0$, resulting in a larger range of $T_{\rm surf}$ capable of hosting liquid water. Some satellites could even become hotter than the boiling temperature of water. \textbf{With increasing $p_0$}, we also notice a clear trend of increasing time spent in the HZ, as it is also observed in Fig.\,\ref{final_plot}.

In Fig.~\ref{pdf} we report, \textbf{above} the hatched area, the probability for a moon to lie in the HZ during the entire temporal evolution, calculated as the integral over the bins in the HZ.  However, to observe such an object, we should determine the timeframe in which liquid water is present. Liquid water conditions could be reached by the satellites also during their temporal evolution, and not necessarily from the beginning of their tidal evolution. In addition to this, if we assume that liquid water is a crucial ingredient for the emergence of life, this timeframe should be compatible with the timescale of producing basic life forms \citep{Cavalazzi2021}. 

In Fig.\,\ref{final_plot} we report the distribution of the time spent in the HZ by our set of moons. We note that as the surface pressure and the total mass of the atmospheric envelope increase, also the number of moons in the HZ increases. \textbf{For $p_0 = 0.1$~bar, moons could be habitable up to 7.3~Myr,} while for $p_0 = 1$~bar a maximum time of \textbf{52~Myr} is reached and the peak of the distribution is at \textbf{45~Myr}. Going to more massive atmospheres ($p_0 = 10$~bar) the maximum and the peak are found at \textbf{276 and 180~Myr}, respectively. In the $p_0 = 100$~bar case, we notice that an increasing number of moons meets the habitable conditions up to \textbf{1.6~Gyr, with a peak at 750~Myr.}

We note that given the assumptions and the limitations of our model, the probability of developing habitable conditions on the surface is proportional to the atmosphere's surface pressure. However, this parameter alone is not sufficient to determine the presence of liquid water. \textbf{From now on, we will focus only on the $p_0$ = 1, 10, and 100~bar cases, where we have a larger number of moons experiencing habitable conditions. For the lower $p_0$, the small number of habitable moons implies a stronger dependence on the initial sample from \citet{cilibrasi2021}, and, for the sake of clarity, the few HZ moons are only reported in Fig.~\ref{appC} in Appendix~C.}

In Fig.~\ref{e_vs_a} we show the dependence of reaching the HZ on the orbital parameters of the moons. The temporal evolution of the different pressure cases is shown. In grey, we represent the sub-HZ satellites (i.e., colder than the HZ conditions), in orange the super-HZ ones (i.e., hotter than the HZ conditions), while the blue satellites lie in the HZ. We note that the number of moons in the HZ increases when increasing the surface pressure, as already observed in Fig.\,\ref{final_plot}. The super-HZ satellites are usually the ones closer to the FFP, and they travel over the HZ, during the orbital circularization, before getting colder than the freezing temperature of water. In Fig.\,\ref{e_vs_a}, we also observe the temporal evolution of the orbital parameters, in particular the decay of the eccentricities \textbf{due to tidal dissipation}. \textbf{It also shows that only the moons with a relatively high eccentricity could reach habitable conditions. This effect is reduced by increasing the surface pressure, where moons with smaller eccentricities could still be found in the HZ}. All the HZ and super-HZ moons reside in a cluster found for \textbf{close-in and eccentric} satellites. \textbf{In particular, the moons initially closer to the planet and that migrate less, spend a large fraction of their evolution in the HZ. This is shown in Fig.\,\ref{e_vs_a}, where these moons populate the HZ cluster as it shrinks in time.}

While in Fig.\,\ref{final_plot} we show how long moons lie in the HZ, \textbf{in Fig.~\ref{e_vs_a}} we focus on when the habitable conditions are met during the tidal evolution. For $p_0$ = 1~bar, habitable moons disappear after 10~Myr (note the logarithmic scale of the times), and their number remains relatively stable before that \textbf{(approximately \textbf{400})}. For $p_0$ = 10~bar, moons stay in the HZ up to \textbf{100~Myr}, and super-HZ moons disappear after \textbf{10~Myr}, partially populating the remaining HZ moons in the next timeframe. Lastly, for $p_0$ = 100~bar, HZ moons are still present at \textbf{1~Gyr}, while super-HZ moons disappear only after \textbf{100~Myr}. As a general trend, the number of HZ moons increases with increasing $p_0$. In Appendix~C, the temperature of the moons in the HZ is reported with an additional figure.

\begin{figure*}
    \centering
    \includegraphics[width=\textwidth]{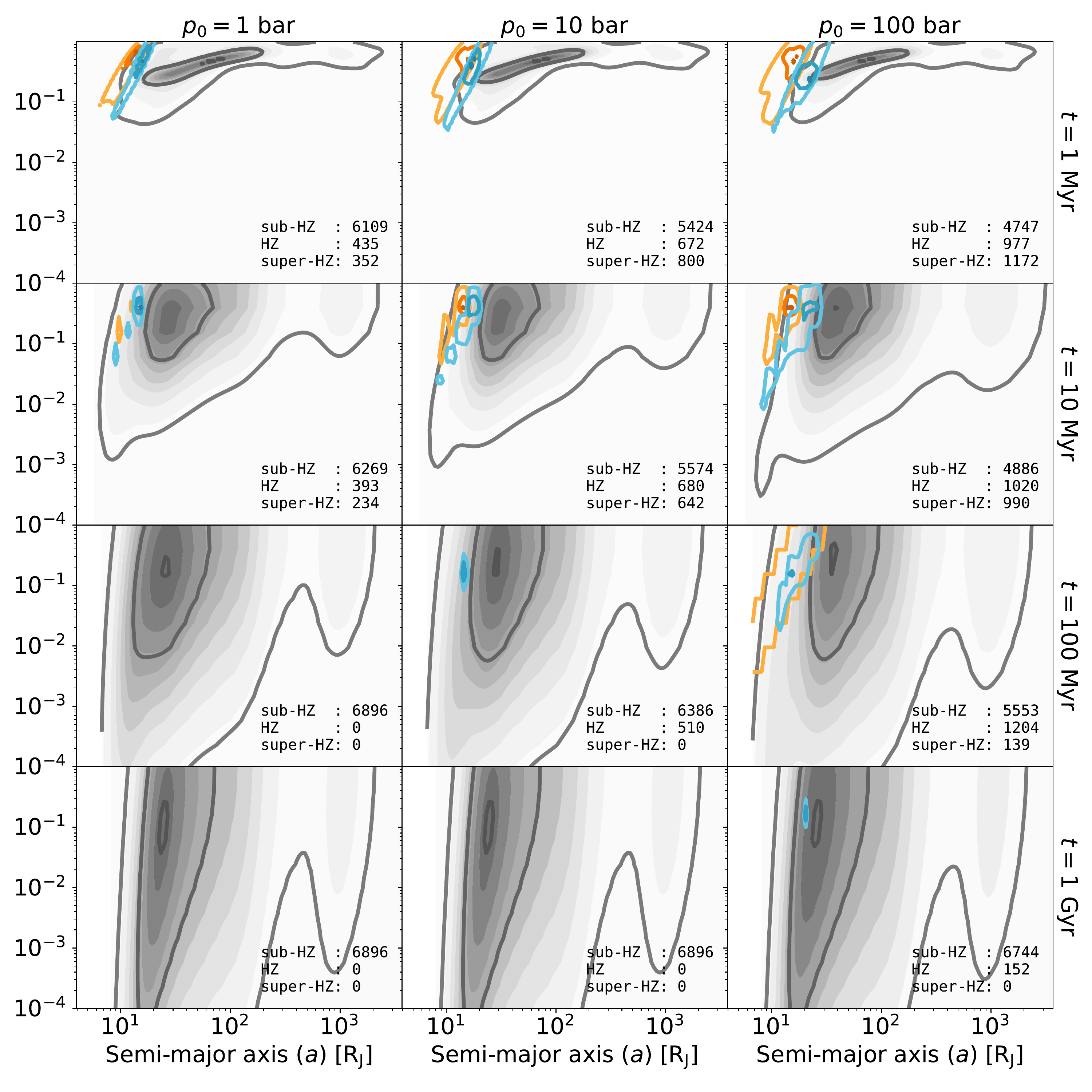}
    \caption{Eccentricity and semi-major axis of Earth-mass moons which survived Sim1, do not enter the Roche radius of the FFP, and can retain an atmosphere. The temporal evolution of the moons' parameters is shown in the vertical direction, while different surface pressure conditions are represented in the different columns. The moons are divided as follows: sub-HZ moons are the ones with a surface temperature colder than the freezing point of water, super-HZ are the warmer ones (above the boiling point), while HZ moons are capable of retaining liquid water on their surfaces. The sub-HZ moons are represented with the KDE and the grey contour lines, which show the 5th, 50th, and 95th percentiles. With the same percentiles, also HZ and super-HZ moons are represented with a KDE in blue and orange respectively. We note that the number of moons in the HZ increases for higher surface pressures, and satellites with larger $p_0$ can remain habitable for longer times. In Appendix~C the surface temperatures of the HZ moons are explicitly shown. The normalization of the KDE is analogous to \fig{KDE}.}
    \label{e_vs_a}
\end{figure*}

From Eq.\,\ref{eq:tidal_power_law}, we observe that the semi-major axis plays the most important role in determining which moons are habitable. Due to the tidal interaction between the moon and the FFP, the energy lost heating up the surface of the moon determines its orbital decay. In this way, some eccentric satellites, which have an initial semi-major axis larger than 10-20 R$_{\rm J}$ (depending on $p_0$), migrate to closer orbits, reaching habitable conditions at a later time. \textbf{However, these migrating satellites quickly circularize their orbits, leaving the HZ at a relatively early timescale.} On the other hand, the more distant, less eccentric satellites are colder and experience less energy loss from the tidal heating mechanism, making their orbits relatively more stable during their temporal evolution. 
\textbf{As expected from Eq.\,\ref{eq:tidal_power_law}, the initial semi-major axis $a_0$ in the tidal model is crucial in determining the timescale for the circularization of the orbit, and thus to keep the tidal heating mechanism effective. High initial orbital eccentricities $e_0$ make distant moons potentially habitable, since the HZ region shifts to larger orbital configurations for high eccentricities (see Fig.\,\ref{appC}).}

\subsection{Non-uniform mass distribution}

\textbf{As reported in Eq.\, \ref{eq:tidal_power_law}, the mass of the moons, along with the orbital parameters, determines their tidal evolution. In order to take into account this aspect, in addition to the constant Earth-mass model, we study the habitability of the moons with the mass distribution from \citet{cilibrasi2021}. It initially includes relatively low-mass satellites, and, after the ejection, only 8\% of moons are more massive than Europa. We repeat the same analysis as of the Earth-mass case using the different atmospheric surface pressures. In this case, the total mass of the atmospheric envelope of each satellite varies as its maximum altitude changes with the surface gravity and the effective temperature (Eq.\,\ref{zmax}). In addition to this, we limit our calculations to $p_0\geq 1$~bar, since with lower values very few moons enter the HZ. Low-mass satellites are not expected to retain dense atmospheres (i.e., with $p_0$ = 100~bar), as observed in the Solar System. }

\textbf{However, it is worth mentioning that, even with  these limiting assumptions, the low-mass satellites do not generate enough tidal heating to heat up the atmosphere and their surface, and to produce liquid water. Another problem that could affect this class of moons is the atmospheric loss via thermal escape; however, they are too cold to make this process effective, and only the moons that enter the Roche limit will produce enough heating, but these are removed from our sample by construction.}

\begin{figure*}
    \centering
    \includegraphics[width=\textwidth]{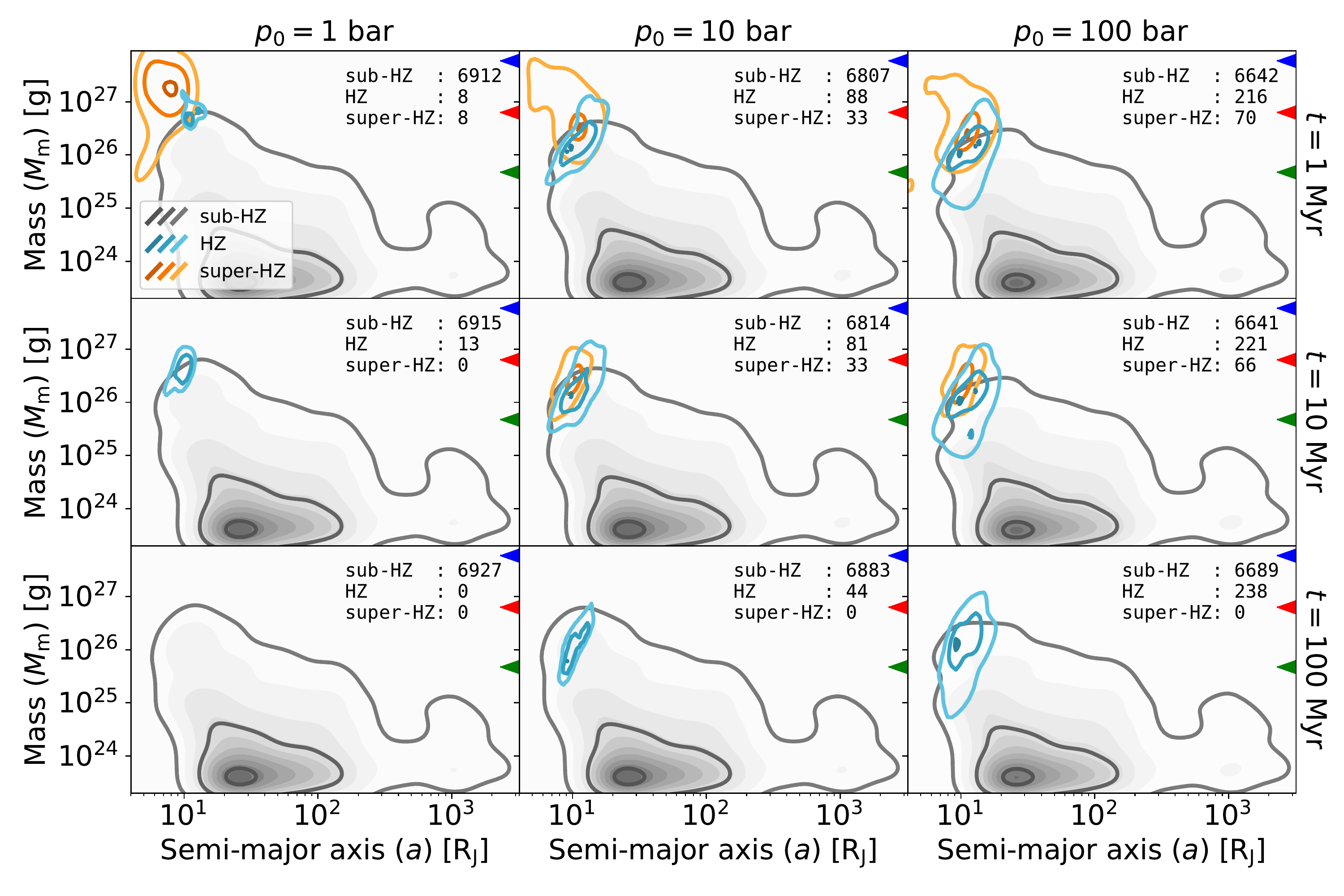}
    \caption{Mass and semi-major axis of moons with \citet{cilibrasi2021}'s masses which survived Sim1, do not enter the Roche radius of the FFP, and can retain an atmosphere. The temporal evolution of the moons' parameters is shown in the vertical direction, while different surface pressure conditions are represented in the different columns. The moons are divided as follows: sub-HZ moons are the ones with a surface temperature colder than the freezing point of water, super-HZ are the warmer ones (above the boiling point), while HZ moons are capable of retaining liquid water on their surfaces. The sub-HZ moons are represented with the KDE and the grey contour lines, which show the 5th, 50th, and 95th percentiles. With the same percentiles, also HZ and super-HZ moons are represented with a KDE in blue and orange respectively. We note that the number of moons in the HZ increases for higher surface pressures, and satellites with larger $p_0$ can remain habitable for longer times. Less massive satellites could become habitable with more substantial atmospheres. \textbf{The triangles show the mass of the Earth (in blue), Mars (in red), and Europa (in green) for comparison}. The normalization of the KDE is analogous to \fig{KDE}.}
    \label{cilibrasi_mass}
\end{figure*}

\begin{figure*}
    \centering
    \includegraphics[width=\textwidth]{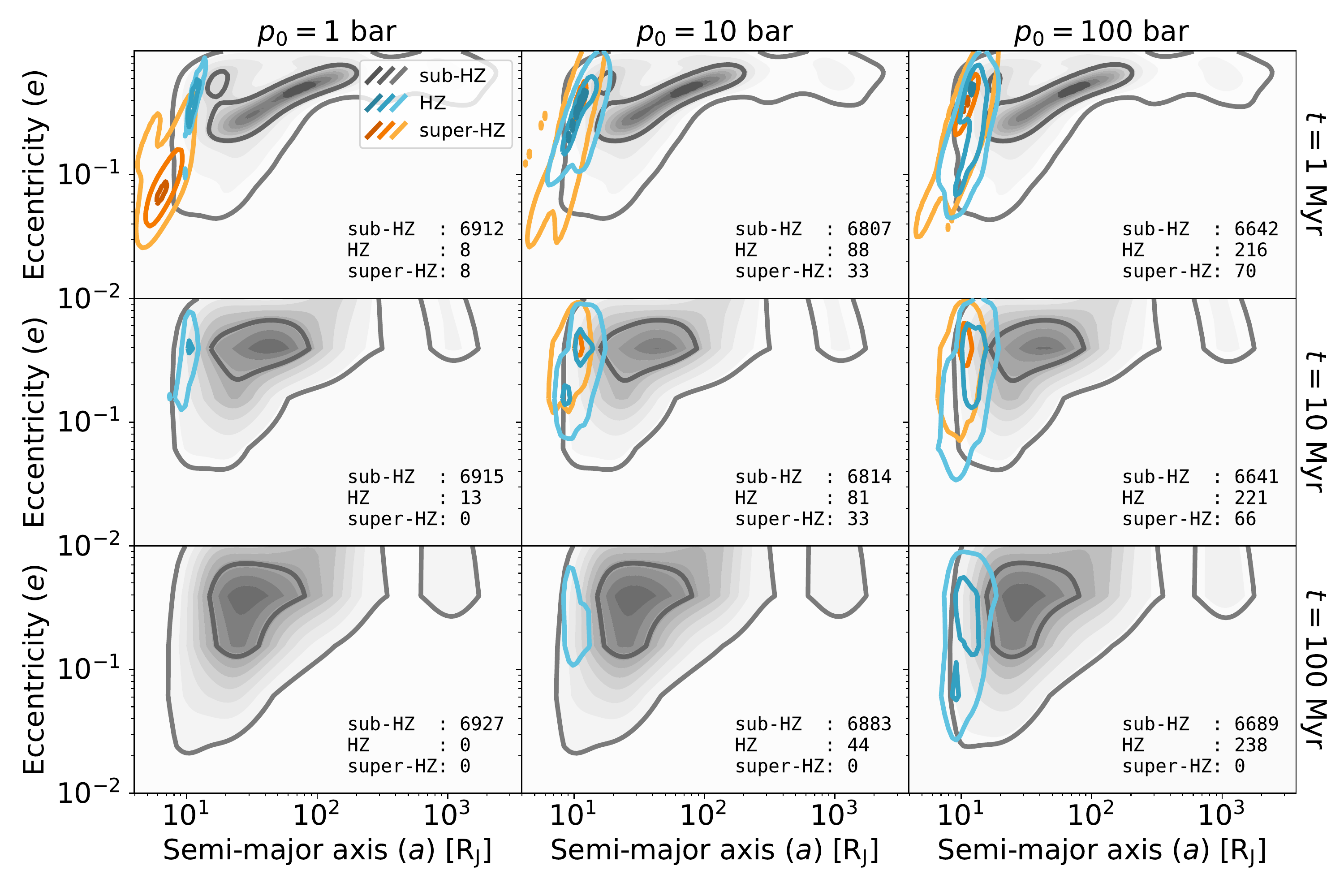}
    \caption{Same as Fig. \ref{e_vs_a}, but for moons whose masses are taken from the \citet{cilibrasi2021} catalogue.}
    \label{cilibrasi_e}
\end{figure*}

\textbf{In Figs. \ref{cilibrasi_mass} and \ref{cilibrasi_e}, we show the influence of the mass, semi-major axis, and eccentricity on the number of habitable moons and on their presence in the HZ.  Fig.~\ref{cilibrasi_mass} shows that only the more massive satellites populate the HZ. Denser atmospheres decrease the minimum mass of habitable satellites. In particular, at $p_0$ = 1~bar the minimum mass of the satellites in the HZ is $M_{\rm m}$ = $1.81 \cdot 10^{25}$~g, at $p_0$ = 10~bar is $M_{\rm m}$ = $4.76 \cdot 10^{24}$~g, and at $p_0$ = 100~bar is $M_{\rm m}$ = $2.21 \cdot 10^{24}$~g. Regarding the orbital parameters, we observe the same general trend as for the Eart-mass case (Fig.~\ref{e_vs_a}), but to enter the HZ, the moons should reach smaller and more eccentric orbital configurations.}

\textbf{In Fig. \ref{cilibrasi_e} we also observe a less pronounced evolution of the orbital parameters (in particular the eccentricity) for the sub-HZ moons, compared to Fig. \ref{e_vs_a}. This is because, very low-mass satellites, which also constitute the majority of the moons, do not strongly tidally interact with the FFP. This translates to more stable orbits during the temporal evolution, as well as less substantial surface heating coming from tidal dissipation. With respect to the orbital parameters, the mass plays a minor role in tidal evolution. For instance, the most massive moon of \citet{cilibrasi2021}'s sample ($M_{\rm m} = 5.42 \cdot 10^{27}$~g), even with $a_0$ = 5.06~R$_{\rm J}$, but with a small eccentricity ($e_0$ = 0.04), experiences a particularly fast orbital circularization ($\lesssim$ 1~Myr), resulting in a short time spent in the HZ.}

\textbf{Assuming this mass distribution, the moons will populate the HZ for timescales shorter than 1~Gyr. This is because, given the smaller number of massive satellites in the distribution, it is statistically less probable to find a moon with orbital parameters compatible with long HZ timescales. In fact, for $p_0$ = 1~bar, moons could populate the HZ up to 15~Myr, while for $p_0$ = 10~bar and $p_0$ = 100~bar, habitable conditions are retained up to 166~Myr and 917~Myr, respectively.}


\subsection{Limitations}

Our results are biased by the many assumptions made throughout the modelling. First, we assumed a planetary system with three Jupiter-like planets. Dynamical simulations with different masses of the perturbers were performed by \citet{hong+et+al2018}, showing that more (less) massive perturbers have smaller (larger) stability boundaries. Regarding the number of giant planets per system, increasing the number of simulated planets also increases the number of ejected FFPs from the system \citep{anderson2019}. Changing this number could affect the dynamics and thus the survivability rate of the moons.

Regarding the survivability of the moons around the escaping planet, we only studied the effect of a last strong close encounter on their orbital parameters, completely neglecting the role played by previous close encounters on the initial population and orbital parameters of the moons. Studying more in detail a single simulation (Sim1) with a considerably small last close encounter, the more distant moons were anyway lost during the evolution, and their orbital parameters were substantially affected also not considering the entire dynamical evolution. Even moons captured by the ejecting planet from the perturber are potentially capable of ending up in very eccentric orbits, which \textbf{could allow distant satellites to become habitable.}

Since our moons are considered massless particles during the dynamical evolution, their mutual interaction was not calculated. Also after the ejection process, in the tidal model, we assumed only single planet-moon systems, completely neglecting the interaction among different satellites in the tidal evolution. Migrating satellites, under the effect of tidal dissipation, are likely captured in resonant configurations, which can further increase the timescale of the tidal heating mechanism. \citet{rabago+steffen2019} demonstrated that both mean-motion and Laplace resonant configurations of satellites placed at the Galilean system orbital distances could also survive the ejection process in most of the simulations. This suggests that previously resonant-captured satellites are possible to survive the dynamical scattering event.

\textbf{In our model, we consider Earth-mass satellites orbiting around a Jupiter-mass FFP. Although such a massive satellite is not present in the Solar System, according to population synthesis studies, e.g., \citet{cilibrasi2018} and \citet{cilibrasi2021}, Earth-mass moons could form around a Jupiter-mass planet, even though they are not the most probable outcome, at least following their findings. However, as of today, there are no observational constraints to the mass of such objects. More massive FFPs are also expected to be able to form and retain more massive moons, but this is beyond the setup of our work. The moons that follow the mass distribution of \citet{cilibrasi2021} are dominated by very low-mass objects, i.e., much smaller than the Galilean moons. These low-mass satellites are introduced in the initial distribution as satellite seeds that are continuously injected during the simulation; at the later stages of the evolution, the almost not-accreted seeds impact the final mass distribution.}

In general, population synthesis studies of forming satellites around giant planets are still very challenging, and a difference in their initial statistics can affect our final results. However, we expect different initial statistics to impact more on the background grey distribution in Figs.\,\ref{e_vs_a}, \ref{cilibrasi_mass} and~\ref{cilibrasi_e} than on the HZ and super-HZ distributions. We expect the close encounter to act on the moons in the way described in Figs.\,\ref{fig:sim1} and~\ref{fig:sim2} and the subsequent tidal evolution to heat up the moons in the cluster identified in Figs.\,\ref{e_vs_a}, \ref{cilibrasi_mass} and~\ref{cilibrasi_e}. 

The atmosphere model we implemented considers a CO$_2$-dominated atmosphere, as in general, an optically-tick atmosphere plays an important role in trapping the heat generated by tidal friction. The composition of a hypothetical exomoon atmosphere is still very uncertain, both for lack of observations, as well as because the composition is probably linked with the formation history of the system. Less substantial atmospheres with respect to the ones considered in \textbf{Fig. \ref{e_vs_a}} are probably more common, affecting our final results and restricting the parameter space and timescale in which habitable conditions can be sustained by the moons, \textbf{as shown in Fig. \ref{pdf} and Fig. \ref{appC} (Appendix C)}. Nevertheless, more massive satellites should be able to form around more massive giant planets and could be potentially considered better candidates to retain more massive atmospheric envelopes.


\section{Discussion and Conclusion}

Liquid water is believed to be a crucial ingredient for the emergence of life as we know it. It could be found on the surface of an exoplanet if the semi-major axis of the planet is within the HZ of its host star. However, it is believed that other objects, like the exomoons orbiting giant FFPs, might retain oceans of liquid water, even if they do not lie in the canonical definition of stellar HZ \citep{avila2021}. FFPs could be the result of the early ejection of a giant planet from its stellar system, and exomoons formed and orbiting the FFP might survive the ejection process remaining bound to it \citep{hong+et+al2018, rabago+steffen2019}. On such exomoons, the presence of liquid water is made possible by tidal and radiogenic heating, together with the presence of an optically thick atmosphere. However, the timescale for this scenario is controlled by the evolution of the orbital parameters, which are first influenced by the ejection process and then by the tidal evolution. The goal of this study is to investigate the potential timescale for the presence of liquid water on the surfaces of these exomoons, and which are the initial orbital parameters, masses, and atmospheric conditions which allow habitable conditions.

To this aim, we performed a set of dynamical simulations of the ejection process of a Jupiter-like planet from a planetary system with three giant planets around a Sun-like star, and we studied the effect of the last close encounter on the stability and survivability of the moons. To model our statistical analysis, we rely on the initial parameter distribution from the population synthesis model of \citet{cilibrasi2021}, where orbital parameters (and masses) for 26\,293 moons are provided. We can summarize our main findings of the dynamical evolution, and in particular on the effect of the last close encounter on the orbits of the moons, as follows:
\begin{itemize}
    \item the orbits of the surviving moons are strongly affected by the last close encounter;
    \item analogously, the impact parameter of the last close encounter plays a major role in the survivability rate of the moons, as already shown by \citet{hong+et+al2018};
    \item \textbf{surviving} moons reach relatively more eccentric and distant configurations, with the outer moons being more affected by the dynamic process.
\end{itemize}

We then selected a simulation with a considerably small impact parameter ($b = 15.36\,{\rm R_J}$), and we evolved the orbits of the \textbf{surviving} moons with a tidal model \citep{hut1981, bolmont2011}. In this way, we could determine when the circularization happens, i.e., when the tidal heating becomes negligible. The time-dependent tidal heating mechanism is also coupled with a CO$_2$-dominated atmosphere, in order to compute the temperature at the surface of the moons. We found that:
\begin{itemize}
    \item \textbf{close-in and eccentric moons are the best candidates for habitable conditions, but moons with very small orbits could exceed the boiling point of water (see Figs.~\ref{e_vs_a} and~\ref{appC});}
    \item moons that experience substantial tidal heating migrate to closer orbits, while starting to circularize. In this scenario, moons can enter the HZ even later in the temporal evolution, when they reach closer orbits;
    \item when the circularization happens ($e\to0$), moons stop their inward migration and their tidal energy drops to zero. 
\end{itemize}

Despite during their lifetime several of the modelled moons present conditions that allow the presence of liquid water, they need to be maintained for a timescale compatible with life to emerge. Regarding the habitability conditions, our main findings are:
\begin{itemize}
    \item \textbf{moons with a relatively small initial semi-major axis after the ejection ($a \sim 15-25$~R$_{\rm J}$) are capable of retaining habitable conditions for long timescale, depending on the surface pressure assumed;}
    \item \textbf{stable orbital eccentricities are needed for long timescales in the HZ, even though the initial orbital eccentricity in the tidal model is not as crucial as the initial semi-major axis;}
    \item an increase in the mass and the density of the atmosphere (i.e., moving to larger surface pressures) significantly increases the number of moons found in the HZ, as well as their timeframe on retaining habitable conditions (see Figs. \ref{final_plot} and \ref{e_vs_a});
    \item moons with a surface pressure of $p_0 = 1$~bar have a maximum time spent in the HZ of 52~Myr, for $p_0 = 10$~bar of 276~Myr and for $p_0 = 100$~bar of 1.60~Gyr.
\end{itemize}

\textbf{We conclude that the moons closer to the hosting planet are the most likely to remain bound during the ejection from the planetary system. These close-in moons are the most suitable for retaining habitable conditions for the longest timescale, which, when assuming denser atmospheres, are comparable with the emergence of life. In particular, for Earth-mass moons with Venus-like atmospheric conditions, around 2\% of them could be found in the HZ beyond 1~Gyr, which is comparable to the timescale of the emergence of life on Earth \citep{Cavalazzi2021}.}

The results of our study represent a lower estimate of the number of habitable moons orbiting a FFP. First, we considered the tidal heating mechanism as the only energy source, but other heating mechanisms could significantly contribute, at different scales, to increase the total energy budget of the moons. For example, radiogenic heating could play a key role in the total energy budget, depending on the internal composition and the history of the planetary system \citep{frank2014}. \textbf{The assumption of a CO$_2$-dominated atmosphere represents a special case that favours habitability, being CO$_2$ an effective greenhouse gas. In addition, we also considered relatively massive atmospheres, which might be less realistic for lighter satellites. However, even under these limiting conditions, the least massive moons do not reach habitable conditions, indicating that massive moons are needed for tidal heating to be effective.}

\textbf{It should also be noted that, at pressures of 0.1~bar, wet-dry cycles, which are considered a promising pathway for the polymerization of RNA \citep{ianeselli2022march}, leads to faster dryings at lower temperatures; This might enhance the polymerization of RNA \citep{dass2022}, while still providing enough amount of CO$_2$ to boost the strand separation due to pH and salt cycling \citep{ianeselli2022jan, ianeselli2022march}.}

We also neglected the resonance configurations that could potentially originate among moons in the tidal evolution. \citet{rabago+steffen2019} demonstrated that resonant configurations could be preserved during the ejection process of the FFP. These resonant configurations might help maintain the orbit eccentric, hence leading to a longer survival timescale of the tidal heating mechanism, as seen for example in the Galilean system \citep{yorder1979}. We did not investigate the timescale of the tidal heating mechanism on moons with resonant configurations, but we expect them to increase the final statistics of habitable moons. However, additional simulations are required.

FFPs could also result from other possible mechanisms, which were not investigated in this work. We can extend our definition of FFPs by including brown dwarfs (BDs), which are supposed to form with a similar mechanism as a planetary system, not reaching the 13\,M$_{\rm J}$ threshold. In this scenario, a central BD could form and retain moons (or planets). Without an ejection process, the potential moons orbiting around such a BD will probably have more circular orbits. \textbf{Given the importance of the orbital eccentricity on tidal heating, secondary objects formed in situ around a BD should have smaller orbit to maintain habitable conditions periods longer than a few Gyr.}

One of the major findings of our work is that exomoons with optically thick atmospheres, with $p_0 = 100$~bar, could retain liquid water on their surfaces \textbf{for billions of years, which is a timescale compatible with the emergence of life}. The composition of the atmosphere itself, the mass of the envelope (and thus the surface pressure), and its evolution in time play a very important role in determining the amount of water that can be produced \citep{avila2021} and its timescale of survivability. More precise models that couple the internal structure and evolution of the mantle and the crust of these moons are needed to better constrain the atmospheric conditions.

\citet{limbach2021} analysed 57 known FFPs and the possibility of observing exomoons eventually orbiting around them. They found that the transit of exomoons on these objects should be frequent, with moons' orbital period of only a few days, and detectable with a transit depth of around 0.1\%-2\%, with bright FFPs representing the best targets to be observed. The JWST will observe WISE~0855 in Cycle~1 for 11~hours and the observation will be sensitive to exomoons as small as Titan/Ganymede with $5 \sigma$. From our results, assuming a close-in ($a \lesssim 15\,{\rm R_J}$) and eccentric orbit, the potential moon can retain liquid water on its surface also considering only a surface pressure for a CO$_2$-dominated atmosphere of 10~bar, with bigger objects being even more favourable candidates, also depending on the mass of the FFP.

\section{Acknowledgements}
\textbf{We thank the Referee who greatly improved the quality of our work.} We thank M.~Cilibrasi and J.~Szul{\'{a}}gyi for providing the electronic version of the data of their previous work. We also want to thank M.~Sterzik for the helpful discussion. This research was supported by the Excellence Cluster ORIGINS, which is funded by the Deutsche Forschungsgemeinschaft (DFG, German Research Foundation) under Germany's Excellence Strategy – EXC-2094 – 390783311 (\url{http://www.universe-cluster.de/}).

\section{Conflict of interest}
The authors declare that they have no conflict of interest.

\section{Appendix A}

The polynomial equations to solve the tidal model (Eqs.\\ref{eq:e}, \ref{eq:e}, \ref{eq:omega_m}, and \ref{eq:omega_p}) are 
\begin{align*}
    N_{\rm a1}(e) = \frac{1 + 31/2e^2 + 255/8e^4 + 185/16e^6 + 85/64e^8}{(1 - e^2)^{15/2}}
\end{align*}

\begin{align*}
    N_{\rm a2}(e) = \frac{1 + 15/2e^2 + 45/8e^4 + 5/16e^6}{(1- e^2)^6}
\end{align*}

\begin{align*}
    N_{\rm e1}(e) = \frac{1 + 15/4e^2 + 15/8e^4 + 5/64e^6}{(1- e^2)^{13/2}}
\end{align*}

\begin{align*}
    N_{\rm e2}(e) = \frac{1 + 3/2e^2 + 1/8e^4}{(1- e^2)^5}
\end{align*}

\begin{align*}
    N_{\rm o1}(e) = \frac{1 + 15/2e^2 + 45/8e^4 + 5/16e^6}{(1- e^2)^{13/2}}
\end{align*}

\begin{align*}
    N_{\rm o2}(e) = \frac{1 + 3e^2 + 3/8e^4}{(1- e^2)^5} \,,
\end{align*}
and are taken from \citet{bolmont2011}.

\section{Appendix B}

In Eq.\,\ref{eq:omega_p}, the temporal evolution of the planetary radius plays the most dominant role in determining the spin evolution of the planet, and thus the evolution of the orbital parameters. 

\begin{figure}[h]
    \centering
    \includegraphics[width=0.5\textwidth]{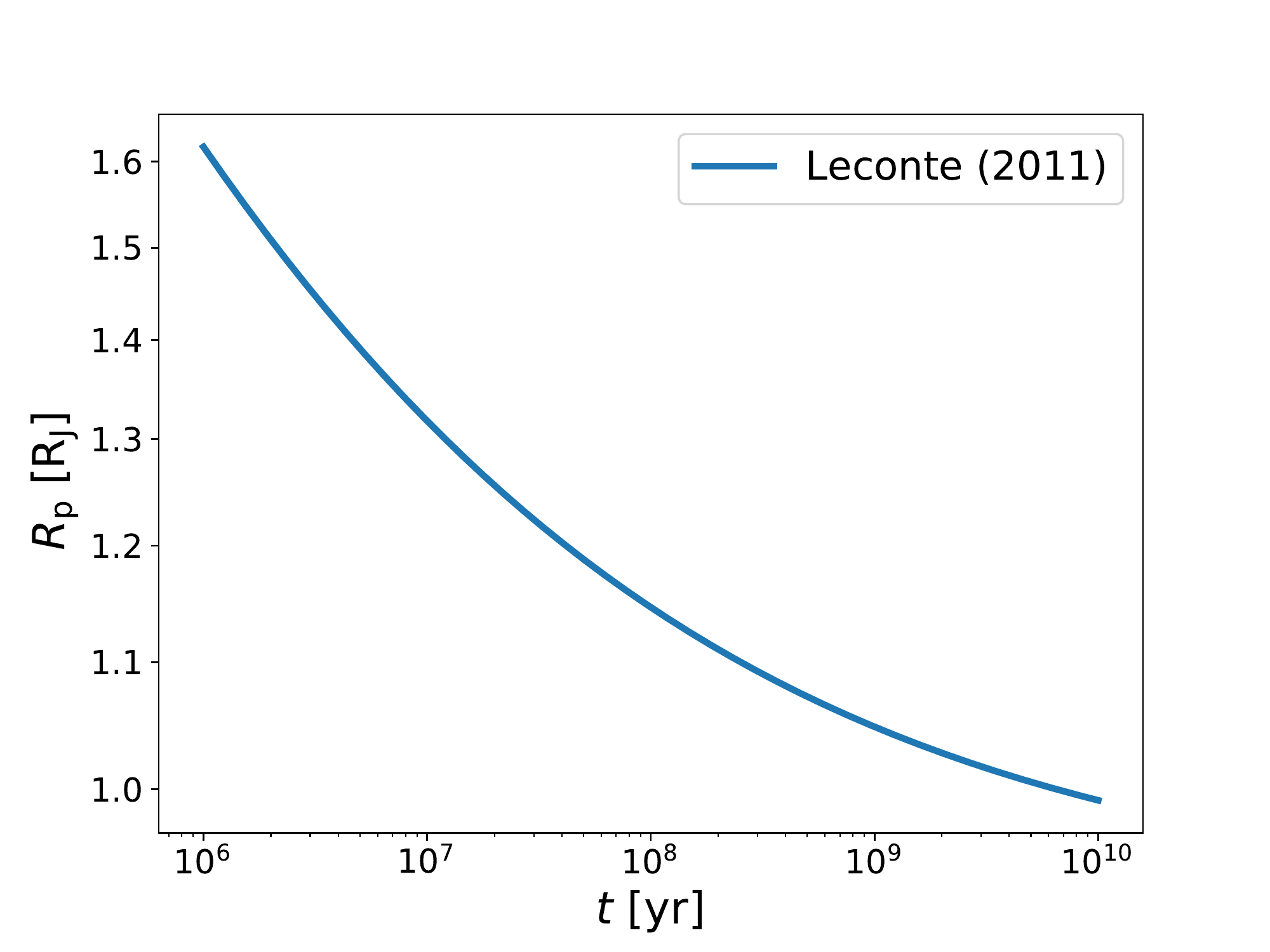}
    \caption{Temporal evolution of the planetary radius following the assumption for a FFP with a Jupiter-mass from \citet{leconte2011}.}
    \label{appendix-B}
\end{figure}

\textbf{The planetary radius evolution profile is taken from \citet{leconte2011}, for a Jupiter-mass FFP that shrinks from 1.6\,R$_{\rm J}$ to 1.0\,R$_{\rm J}$, as shown in Fig.~\ref{appendix-B}. Their profile is modelled as}
\begin{equation}
    f = A \cdot e^{-B \log_{10}(t)} + C\,,
\end{equation}
\textbf{with $A$ = 20.043 R$_{\rm J}$, $B$ = 0.559, and $C$ = 0.917 R$_{\rm J}$}. 

\textbf{The profile used for the radius evolution of the planet plays a major role in the tidal model. The factor ${\dd R_p}/{\dd t}$ dominates Eq.\,\ref{eq:omega_p}, and thus the planetary spin depends on the selected profile. Since the tidal interaction between the planet and the moon depends on the position of the tidal bulge, a difference in planetary spin will result in a different evolution of the orbital parameters. }

\section{Appendix C}

\begin{figure*}
    \centering
    \includegraphics[width=\textwidth]{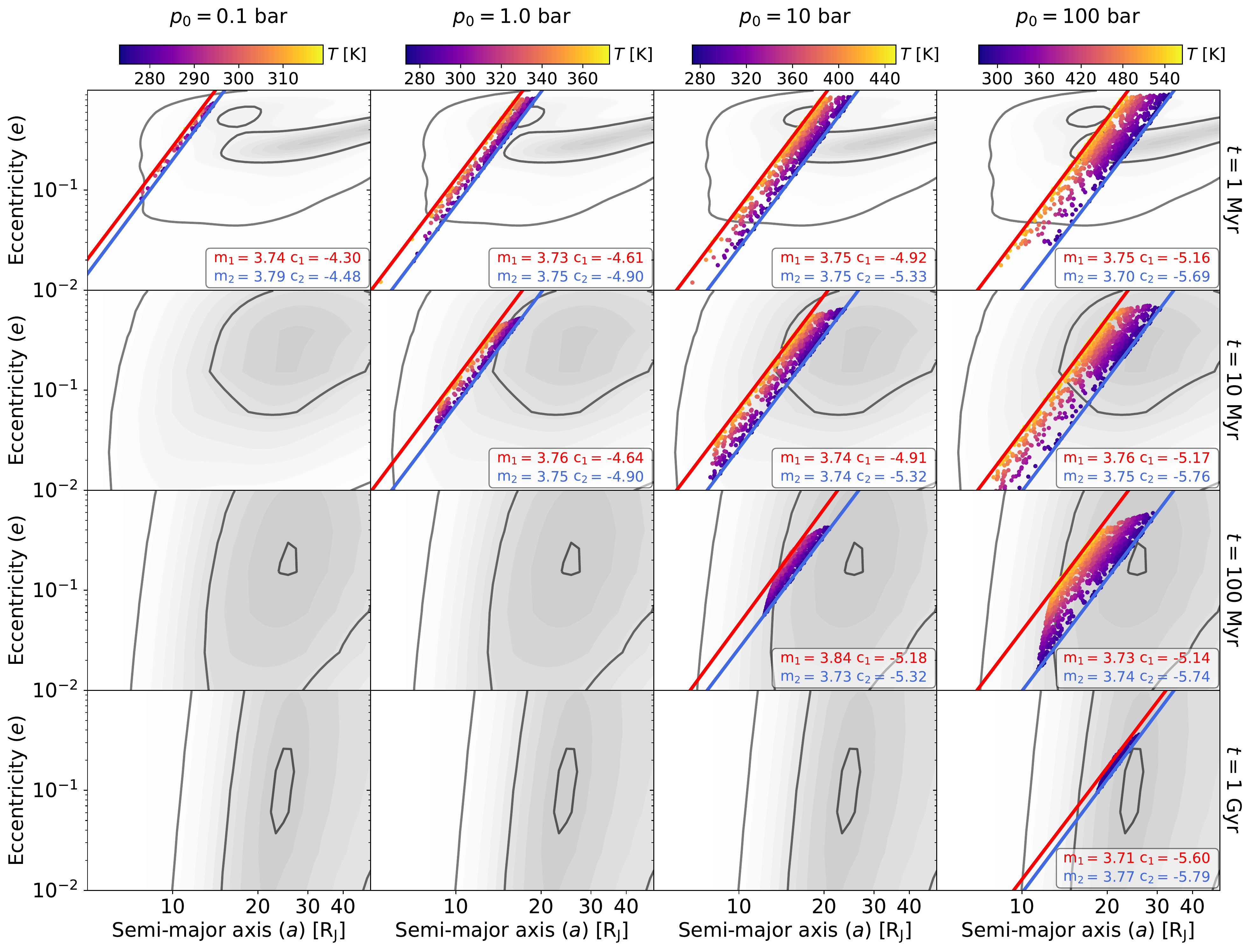}
    \caption{Same as Fig. \ref{e_vs_a}, with a focus on the HZ cluster. The grey contour lines represent all the moons in the investigated population, while the colour dots indicate the surface temperature of the HZ satellites. With the red and blue solid lines, we fit the HZ cluster boundaries, and the fit parameters are reported in each panel with habitable moons. The colour bar indicates the surface temperature between freezing and boiling point, and it changes increasing the surface pressure of the atmosphere.}
    \label{appC}
\end{figure*}

\textbf{Fig.\,\ref{appC} shows the surface temperature of the moons in the HZ of Fig.\,\ref{e_vs_a} during the tidal evolution. With respect to Fig.\,\ref{e_vs_a}, here we limit the semi-major axis between 5 and 50~R$_{\rm J}$, and the eccentricity between 0.01 and 1, to focus on the HZ cluster. We also indicate the few habitable moons in the $p_0$ = 0.1~bar. The grey contours are the KDE of all the moons (sub-HZ, HZ, and super-HZ), while the coloured dots are only the HZ satellites. The colour bar for the surface temperatures varies for different $p_0$, since the boiling temperature of water depends on the surface pressure. Again, we observe the clear trend of an increasing number of moons in the HZ, and the correlation between time spent in the HZ and surface pressure.}

\textbf{Fig.\,\ref{appC} reports the correlation between the eccentricity and semi-major axis for the moons that populate the HZ. We fit the boundaries of the HZ region, selected as the 5\% hottest and coldest moons. The red solid line and the blue line indicate respectively the hottest and coldest boundaries. In each panel with habitable moons, we report the parameters of the fit as}
\begin{equation}
    10^{c_2} a^{m_2} < e < 10^{c_1} a^{m_1}\,,
\end{equation}
\textbf{where $a$ is in units of R$_{\rm J}$. The HZ region becomes larger as we move to more massive atmospheres because the range of habitable surface temperature broadens.}

\newcommand{\aj}{Astronomical Journal}
\newcommand{\araa}{Annual Review of Astronomy and Astrophysics}
\newcommand{\apj}{Astrophysical Journal}
\newcommand{\apjl}{Astrophysical Journal, Letters}
\newcommand{\apjs}{Astrophysical Journal, Supplement}
\newcommand{\ao}{Applied Optics}
\newcommand{\apss}{Astrophysics and Space Science}
\newcommand{\aap}{Astronomy and Astrophysics}
\newcommand{\aapr}{Astronomy and Astrophysics Reviews}
\newcommand{\aaps}{Astronomy and Astrophysics, Supplement}
\newcommand{\azh}{Astronomicheskii Zhurnal}
\newcommand{\baas}{Bulletin of the AAS}
\newcommand{\icarus}{Icarus}
\newcommand{\jrasc}{Journal of the RAS of Canada}
\newcommand{\jaavso}{Journal of the American Association of Variable Star Observers}
\newcommand{\memras}{Memoirs of the RAS}
\newcommand{\mnras}{Monthly Notices of the RAS}
\newcommand{\pra}{Physical Review A: General Physics}
\newcommand{\prb}{Physical Review B: Solid State}
\newcommand{\prc}{Physical Review C}
\newcommand{\prd}{Physical Review D}
\newcommand{\pre}{Physical Review E}
\newcommand{\prl}{Physical Review Letters}
\newcommand{\psj}{Planetary Science Journal}
\newcommand{\pasp}{Publications of the ASP}
\newcommand{\pasj}{Publications of the ASJ}
\newcommand{\qjras}{Quarterly Journal of the RAS}
\newcommand{\skytel}{Sky and Telescope}
\newcommand{\solphys}{Solar Physics}
\newcommand{\sovast}{Soviet Astronomy}
\newcommand{\ssr}{Space Science Reviews}
\newcommand{\zap}{Zeitschrift fuer Astrophysik}
\newcommand{\nat}{Nature}
\newcommand{\iaucirc}{IAU Cirulars}
\newcommand{\aplett}{Astrophysics Letters}
\newcommand{\apspr}{Astrophysics Space Physics Research}
\newcommand{\bain}{Bulletin Astronomical Institute of the Netherlands}
\newcommand{\fcp}{Fundamental Cosmic Physics}
\newcommand{\gca}{Geochimica Cosmochimica Acta}
\newcommand{\grl}{Geophysics Research Letters}
\newcommand{\jcp}{Journal of Chemical Physics}
\newcommand{\jgr}{Journal of Geophysics Research}
\newcommand{\jqsrt}{Journal of Quantitative Spectroscopy and Radiative Trasfer}
\newcommand{\memsai}{Mem. Societa Astronomica Italiana}
\newcommand{\nphysa}{Nuclear Physics A}
\newcommand{\physrep}{Physics Reports}
\newcommand{\physscr}{Physica Scripta}
\newcommand{\planss}{Planetary Space Science}
\newcommand{\procspie}{Proceedings of the SPIE}
\newcommand{\actaa}{Acta Astronomica}
\newcommand{\caa}{Chinese Astronomy and Astrophysics}
\newcommand{\cjaa}{Chinese Journal of Astronomy and Astrophysics}
\newcommand{\jcap}{Journal of Cosmology and Astroparticle Physics}
\newcommand{\na}{New Astronomy}
\newcommand{\nar}{New Astronomy Review}
\newcommand{\pasa}{Publications of the Astron. Soc. of Australia}
\newcommand{\rmxaa}{Revista Mexicana de Astronomia y Astrofisica}
\newcommand{\maps}{Meteoritics and Planetary Science}
\newcommand{\aas}{American Astronomical Society Meeting Abstracts}
\newcommand{\dps}{American Astronomical Society/Division for Planetary Sciences Meeting Abstracts}

\bibliographystyle{agsm2}
\bibliography{bibliography}

\end{document}